\documentclass[nonacm,screen,anonymous=false]{acmart}
\usepackage{xspace}
\usepackage{tabularx}
\usepackage{alltt}
\usepackage{geometry}
\usepackage{array}
\usepackage{enumitem}
\usepackage{caption}
\usepackage{subcaption}
\usepackage{ragged2e}

\AtBeginDocument{%
  \providecommand\BibTeX{{%
    \normalfont B\kern-0.5em{\scshape i\kern-0.25em b}\kern-0.8em\TeX}}}

\setcopyright{none}
\copyrightyear{2021}
\acmDOI{}

\begin{document}

\newcommand\lucy[1]{{\color{blue}\{\textit{#1}\}$_{lucy}$}}
\newcommand\isabel[1]{{\color{purple}\{\textit{#1}\}$_{Isabel}$}}
\newcommand\jonathan[1]{{\color{brown}\{\textit{#1}\}$_{Jonathan}$}}
\newcommand\jb[1]{{\color{brown}\{\textit{#1}\}$_{Jonathan}$\footnote{\textcolor{red}{NOTE}}}}
\newcommand\dan[1]{{\color{red}\{\textit{#1}\}$_{dan}$}}

\newcommand\todoit[1]{{\color{red}\{TODO: \textit{#1}\}}}
\newcommand\todo{{\color{red}{TODO}}\xspace}
\newcommand\todocite{{\color{red}{CITE}}\xspace}

\newcommand\scially{{SciA11y}\xspace}

\newcommand{\xcompliance}[1]{Adobe-#1 Compliance\xspace}
\newcommand\numchi{3,248\xspace}
\newcommand\percchiaccessibility{2\%\xspace}

\newcommand\numpdfs{11,397\xspace}
\newcommand\percaccessible{2.4\%\xspace}

\newcommand\numtotalpdfs{{\color{red}{XXX}}\xspace}
\newcommand\numattemptedpdfs{{\color{red}{26,824}}\xspace}

\newcommand\numeval{{385}\xspace}
\newcommand\numevalskipped{{137}\xspace}

\newcommand\numusers{six\xspace}

\newcommand\semanticscholar{{\small{\texttt{anon\_corpus}}}\xspace}
\newcommand\allenai{{\small{\texttt{anonymized}}}\xspace}
\newcommand\githublink{\url{anonymous\_url}\xspace}

\newcolumntype{L}[1]{>{\raggedright\let\newline\\\arraybackslash\hspace{0pt}}p{#1}}
\newcolumntype{M}[1]{>{\raggedright\let\newline\\\arraybackslash\hspace{0pt}}m{#1}}

\newcommand{\rulesep}{\unskip\ \textcolor{gray}{\vrule}\ }

\definecolor{darkgreen}{rgb}{0.0, 0.4, 0.13}

\title{Improving the accessibility of scientific documents}
\subtitle{Current state, user needs, and a system solution to enhance scientific PDF accessibility for blind and low vision users}


\author{Lucy Lu Wang}
\authornote{Denotes equal contribution}
\email{lucyw@allenai.org}
\orcid{0000-0001-8752-6635}
\affiliation{%
  \institution{Allen Institute for AI}
  \city{Seattle}
  \state{WA}
  \postcode{98103}
}

\author{Isabel Cachola}
\authornotemark[1]
\authornote{Work done while at the Allen Institute for AI}
\email{icachola@cs.jhu.edu}
\orcid{}
\affiliation{%
  \institution{The Johns Hopkins University}
  \city{Baltimore}
  \state{MD}
  \postcode{21218}
}

\author{Jonathan Bragg}
\email{jbragg@allenai.org}
\orcid{}
\affiliation{%
  \institution{Allen Institute for AI}
  \city{Seattle}
  \state{WA}
  \postcode{98103}
}

\author{Evie Yu-Yen Cheng}
\email{eviec@allenai.org}
\orcid{}
\affiliation{%
  \institution{Allen Institute for AI}
  \city{Seattle}
  \state{WA}
  \postcode{98103}
}

\author{Chelsea Haupt}
\email{chealseah@allenai.org}
\orcid{}
\affiliation{%
  \institution{Allen Institute for AI}
  \city{Seattle}
  \state{WA}
  \postcode{98103}
}

\author{Matt Latzke}
\email{mattl@allenai.org}
\orcid{}
\affiliation{%
  \institution{Allen Institute for AI}
  \city{Seattle}
  \state{WA}
  \postcode{98103}
}

\author{Bailey Kuehl}
\email{baileyk@allenai.org}
\orcid{}
\affiliation{%
  \institution{Allen Institute for AI}
  \city{Seattle}
  \state{WA}
  \postcode{98103}
}

\author{Madeleine van Zuylen}
\email{madeleinev@allenai.org}
\orcid{}
\affiliation{%
  \institution{Allen Institute for AI}
  \city{Seattle}
  \state{WA}
  \postcode{98103}
}

\author{Linda Wagner}
\email{lindaw@allenai.org}
\orcid{}
\affiliation{%
  \institution{Allen Institute for AI}
  \city{Seattle}
  \state{WA}
  \postcode{98103}
}

\author{Daniel S. Weld}
\email{danw@allenai.org}
\orcid{}
\affiliation{%
  \institution{Allen Institute for AI}
  \city{Seattle}
  \state{WA}
  \postcode{98103}
}
\affiliation{%
  \institution{University of Washington}
  \city{Seattle}
  \state{WA}
  \postcode{98103}
}

\renewcommand{\shortauthors}{Wang LL and Cachola I et al}

\begin{abstract}
  The majority of scientific papers are distributed in PDF, which pose challenges for accessibility, especially for blind and low vision (BLV) readers. We characterize the scope of this problem by assessing the accessibility of \numpdfs PDFs published 2010--2019 sampled across various fields of study, finding that only \percaccessible of these PDFs satisfy all of our defined accessibility criteria. We introduce the \scially system to offset some of the issues around inaccessibility. \scially incorporates several machine learning models to extract the content of scientific PDFs and render this content as accessible HTML, with added novel navigational features to support screen reader users. An intrinsic evaluation of extraction quality indicates that the majority of HTML renders (87\%) produced by our system have no or only some readability issues. We perform a qualitative user study to understand the needs of BLV researchers when reading papers, and to assess whether the \scially system could address these needs. We summarize our user study findings into a set of five design recommendations for accessible scientific reader systems. User response to \scially was positive, with all users saying they would be likely to use the system in the future, and some stating that the system, if available, would become their primary workflow. We successfully produce HTML renders for over 12M papers, of which an open access subset of 1.5M are available for browsing at \href{https://scia11y.org/}{scia11y.org}.
\end{abstract}




\maketitle

\section{Introduction}

Scientific literature is most commonly available in the form of PDFs, which pose challenges for accessibility \citep{NielsenPDFStillUnfit, Bigham2016AnUT}. When researchers, students, and other individuals who are blind or low vision (BLV) interact with scientific PDFs through screen readers, the availability of document structure tags, labeled reading order, labeled headers, and image alt-text are necessary to facilitate these interactions. However, these features must be painstakingly added by authors using proprietary software tools, and as a result, are often missing from papers. Low vision or dyslexic readers who interact with PDFs through screen magnification or text-to-speech may also find the complexity of certain academic paper PDF formats challenging, e.g., non-linear layout can interrupt the flow of text in a magnifying tool. Inaccessible paper PDFs can lead to high cognitive overload, frustration, and abandonment of reading for BLV readers. 

Unfortunately, we find that the majority of scientific PDFs lack basic accessibility features. We estimate based on a sample of \numpdfs PDFs from multiple fields of study that only around \percaccessible of paper PDFs released in the last decade satisfy all of the aforementioned accessibility requirements. 
Accessibility challenges for academic PDFs are largely due to three factors: (1) the complexity of the PDF file format, which make it less amenable to certain accessibility features, (2) the dearth of tools, especially non-proprietary tools, for creating accessible PDFs, and (3) the dependency on volunteerism from the community with minimal support or enforcement \citep{Bigham2016AnUT}. The intent of the PDF file format is to support faithful visual representation of a document for printing, a goal that is inherently divergent from that of document representation for the purposes of accessibility. Though some professional organizations like the Association for Computing Machinery (ACM) have encouraged PDF accessibility through standards and writing guidelines,\footnote{\href{https://www.acm.org/publications/authors/submissions}{https://www.acm.org/publications/authors/submissions}} uptake among academic publishers and disciplines more broadly has been limited. 

While policy changes help, the fact remains that most academic PDFs produced today, and historically, are inaccessible, yet remain as the dominant way to read those papers. A long-range solution will necessitate buy-in from multiple stakeholders---publishers, authors, readers, technologists, granting agencies, and the like. But in the interim, there are technological solutions that can be offered as a sort of ``band-aid'' to the problem. We use this paper to offer an in-depth qualitative and quantitative description of the problem as it stands, and to introduce one such technological solution: the \scially system that automatically extracts semantic information from paper PDFs and re-renders this content in the form of an accessible HTML document. Though the process is imperfect and can introduce errors, we demonstrate the ability of the rendered HTMLs to reduce cognitive load and facilitate in-paper navigation and interactions for BLV users. 

The goals and contributions of this paper are three-fold:

\begin{enumerate}
    \item We characterize the state of academic-paper PDF accessibility by estimating the degree of adherence to accessibility criteria for papers published in the last decade (2010--2019), and describe correlations between year, field of study, PDF typesetting software, and PDF accessibility.
    \item We propose an automated approach for extracting the content of academic PDFs and displaying this content in a more accessible HTML document format. We build a prototype that re-renders 12 million PDFs in HTML, and describe the design decisions, features, and quality of the renders (assessed as faithfulness to the source PDF). We perform expert grading of the rendered HTML and report an error analysis. A demo of our system is available at \href{https://scia11y.org/}{scia11y.org}, which makes available 1.5M HTML renders of open access PDFs.
    \item We conduct an exploratory user study with \numusers BLV scholars to better understand the challenges they experience when reading academic papers and how our proposed tool might augment their current workflow. During the study, we ask users to interact with the prototype and offer feedback for its improvement. We perform open coding of interviews to identify existing reading challenges, coping mechanisms, as well as positive and negative responses to prototype features. We summarize the findings of this user study into a set of design recommendations.
\end{enumerate}

Our analysis reveals that PDF accessibility adherence is low across all fields of study. Of the five accessibility criteria we assess, only \percaccessible of the PDFs we assess demonstrate full compliance. Though compliance for several criteria seems to be increasing over time, author awareness and contribution to accessibility remains low, as Alt-text has the lowest compliance of the five criteria at between 5--10\% (Alt-text is the only criterion of the five that \textit{requires} author intervention in all cases using current tools). We also find that typesetting software is strongly associated with accessibility compliance, with LaTeX and publishing software like Arbortext APP producing low compliance PDFs, while Microsoft Word is generally associated with higher compliance.

\begin{figure}[t!]
    \centering
    \includegraphics[width=\textwidth]{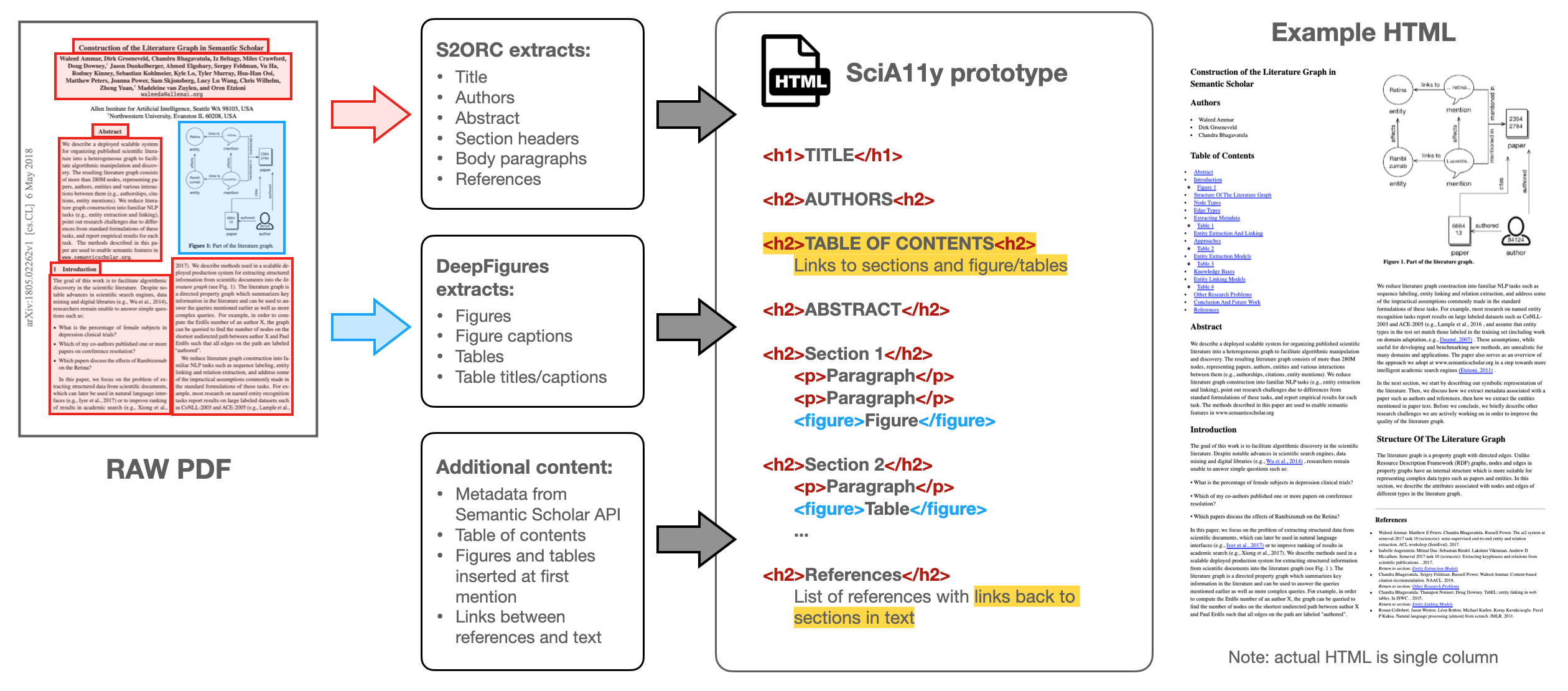}
    \caption{A schematic for creating the \scially HTML render from a paper PDF. Starting with the raw two-column PDF on the left, S2ORC \citep{lo-wang-2020-s2orc} is used to extract title, authors, abstract, section headers, body text, and references. S2ORC also identifies links between inline citations and references to figures and table objects. DeepFigures \citep{Siegel2018ExtractingSF} is used to extract figures and tables, along with their captions. The output of these two models are merged with metadata from the Semantic Scholar API. Heuristics are used to construct a table of contents, to insert figures and tables in the appropriate places in the text, and to repair broken URLs. We add HTML headers as illustrated (header tags for sections, paragraph tags for body text, and figure tags for figures and tables); highlighted components (table of contents and links in references) are not in the PDF and novel navigational features that we introduce to the HTML render. An example HTML render of parts of a paper document is show to the right (actual render is single column, which is split here for presentation).}
    \label{fig:pipeline}
    \Description{A schematic diagram showing the components of the SciA11y pipeline. An image of a paper PDF is on the left. Red boxes on the PDF image highlight the text components from the paper, with an arrow pointing to a box that says "S2ORC extracts: title, authors, abstract, section headers, body paragraphs, and references." A blue box on the PDF image highlights a figure, with an arrow pointing to a box that says "DeepFigures extracts: figures, figure captions, tables, and table titles/captions." A box below "S2ORC extracts" and "DeepFigures extracts" says "Additional content: metadata from Semantic Scholar API, table of contents, figures and tables inserted at first mention, and links between references and text." Arrows from all three boxes point into a larger box that describes the SciA11y prototype, where HTML tags are inserted around various blocks of text extracted from the PDF. On the right of all this is a screen capture of an example HTML render, showing how the semantic content from the PDF is represented as a single-column HTML page for easy reading.}
\end{figure}

To offset the reading challenges of inaccessible papers for BLV researchers, we propose and test the \scially system for rendering academic PDFs into accessible HTML documents. As shown in Figure~\ref{fig:pipeline}, our prototype integrates several machine learning text and vision models to extract the structure and semantic content of papers. The content is represented as an HTML document with headings and links for navigation, figures and tables, as well as other novel features to assist in document structure understanding. Our evaluation of the \scially system identifies common classes of extraction problems, and finds that though many papers exhibit some extraction errors, the majority (55\%) have no major problems that impact readability, and another 32\% have only some problems that impact readability.

Through our user study, we identify numerous challenges faced by BLV users when reading paper PDFs, including some that affect the whole document or limit navigation, and many that affect the ability of the reader to understand text or various elements of a paper like math content or tables. Responses to \scially were positive; participants especially liked navigation features such as headings, the table of contents, and bidirectional links between inline citations and references. Of the extraction errors in \scially, missed or incorrectly extracted headings were the most problematic, as these impact the user's ability to navigate between sections and fully trust the system. All users reported being likely to use the system in the future. When asked how the system might be integrated into their workflow, one participant replied ``I think it would become the workflow.'' Another participant said, ``for unaccessible PDFs, this is life-changing.'' We condense these findings into a set of recommendations for designing and engineering accessible reading systems (Section~\ref{sec:designrecs}). Most importantly, documents should be structured to match a reader's mental model, objects should be properly tagged, and care should be taken to reduce the reader's cognitive load and increase trust in the system. Features that emulate the external memory that visual layout provides to sighted users can be especially beneficial.

This paper is organized as follows. Following a description of related work in Section \ref{sec:related_work}, we first provide a meta-scientific analysis of the current state of academic PDF accessibility in Section \ref{sec:sos}. In Section \ref{sec:pdf2html}, we document our pipeline for converting PDF to HTML and describe the \scially prototype for rendering papers. An evaluation of HTML render quality and faithfulness is provided in Section \ref{sec:evaluation}. Section \ref{sec:user_study} describes our user study and findings. 
We recognize that no PDF extraction system is perfect, and many open research challenges remain in improving these systems. However, based on our findings, we believe \scially can dramatically improve screen reader navigation of most papers compared to PDFs, and is well-positioned to assist BLV researchers with many of their most common reading use cases. Our hope is that a system such as \scially can improve BLV researcher access to the content of academic papers, and that these design recommendations can be leveraged by others to create better, more faithful, and ultimately more usable tools and systems for scholars in the BLV community.

\section{Related work}
\label{sec:related_work}

Accessibility is an essential component of computing, which aims to make technology broadly accessible to as many users as possible, including those with differing sets of abilities. Improvements in usability and accessibility falls to the community, to better understand the needs of users with differing abilities, and to design technologies that play to this spectrum of abilities \citep{Wobbrock2011AbilityBasedDC}.
In computing, significant strides have been made to increase the accessibility of web content. For example, various versions of the Web Content Accessibility Guidelines (WCAG) \citep{Chisholm2001WebCA, Caldwell2008WebCA} and the in-progress working draft for WCAG 3.0,\footnote{\href{https://www.w3.org/TR/wcag-3.0/}{https://www.w3.org/TR/wcag-3.0/}} or standards such as ARIA from the W3C's Web Accessibility Initiative (WAI)\footnote{\href{https://www.w3.org/WAI/standards-guidelines/aria/}{https://www.w3.org/WAI/standards-guidelines/aria/}} have been released and used to guide web accessibility design and implementation. Similarly, positive steps have been made to improve the accessibility of user interfaces and user experience \citep{Peissner2012MyUIGA, Peissner2013UserCI, Thompson2014ImprovingTU, Bigham2014MakingTW}, as well as various types of media content \citep{Mirri2017TowardsAG, Nengroo2017AccessibleI, Gleason2020TwitterAA}. 

We take inspiration from accessibility design principles in our effort to make research publications more accessible to users who are blind and low vision. Blindness and low vision are some of the most common forms of disability, affecting an estimated 3--10\% of Americans depending on how visual impairment is defined \citep{CDCVisionLossBurden}. BLV researchers also make up a representative sample of researchers in the United States and worldwide. A recent Nature editorial pushes the scientific community to better support researchers with visual impairments \citep{NatureCareerColumn2020}, since existing tools and resources can be limited. There are many inherent accessibility challenges to performing research. In this paper, we engage with one of these challenges that affects all domains of study, accessing and reading the content of academic publications. 

BLV users interact with papers using screen readers, braille displays, text-to-speech, and other assistive tools. A WebAIM survey of screen reader users found that the vast majority (75.1\%) of respondents indicate that PDF documents are very or somewhat likely to pose significant accessibility issues.\footnote{\href{https://webaim.org/projects/screenreadersurvey8/}{https://webaim.org/projects/screenreadersurvey8/}} Most paper are published in PDF, which is inherently inaccessible, due in large part to its conflation of visual layout information with semantic content \citep{NielsenPDFStillUnfit, Bigham2016AnUT}. 
\citet{Bigham2016AnUT} describe the historical reasons we use PDF as the standard document format for scientific publications, as well as the barriers the format itself presents to accessibility. Prior work on scientific accessibility have made recommendations for how to make PDFs more accessible \cite{Rajkumar2020PDFAO, Darvishy2018PDFAT}, including greater awareness for what constitutes an accessible PDF and better tooling for generating accessible PDFs. Some work has focused on addressing components of paper accessibility, such as the correct way for screen readers to interpret and read mathematical equations \citep{Flores2010MathMLTA, Bates2010SpokenMU, Sorge2014TowardsMM, Mackowski2017MultimediaPF, Ahmetovic2018AxessibilityAL, Ferreira2004EnhancingTA, Sojka2013AccessibilityII}, describe charts and figures \citep{Elzer2008AccessibleBC, Engel2017TowardsAC, Engel2019SVGPlottAA}, automatically generate figure captions \citep{Chen2019NeuralCG, Qian2020AFS}, or automatically classify the content of figures \citep{Kim2018MultimodalDL}. Other work applicable to all types of PDF documents aims to improve automatic text and layout detection of scanned documents \cite{Nazemi2014PracticalSM} and extract table content \cite{Fan2015TableRD, Rastan2019TEXUSAU}. In this work, we focus on the issue of representing overall document structure, and navigation within that structure. Being able to quickly navigate the contents of a paper through skimming and scanning is an essential reading technique \citep{Maxwell1972SkimmingAS}, which is currently under-supported by PDF documents and PDF readers when reading these documents by screen reader. 

There also exists a variety of automatic and manual tools that assess and fix accessibility compliance issues in PDFs, including the Adobe Acrobat Pro Accessibility Checker\footnote{\href{https://www.adobe.com/accessibility/products/acrobat/using-acrobat-pro-accessibility-checker.html}{https://www.adobe.com/accessibility/products/acrobat/using-acrobat-pro-accessibility-checker.html}}, Common Look\footnote{\href{https://monsido.com/monsido-commonlook-partnership}{https://monsido.com/monsido-commonlook-partnership}}, ABBYY FineReader\footnote{\href{https://pdf.abbyy.com/}{https://pdf.abbyy.com/}}, PAVE\footnote{\href{https://pave-pdf.org/faq.html}{https://pave-pdf.org/faq.html}}, and PDFA Inspector\footnote{\href{https://github.com/pdfae/PDFAInspector}{https://github.com/pdfae/PDFAInspector}}. To our knowledge, PAVE and PDFA Inspector are the only non-proprietary, open-source tools for this purpose. Based on our experiences, however, all of these tools require some degree of human intervention to properly tag a scientific document, and tagging and fixing must be performed for each new version of a PDF, regardless of how minor the change may be.

Guidelines and policy changes have been introduced in the past decade to ameliorate some of the issues around scientific PDF accessibility. Some conferences, such as The ACM CHI Virtual Conference on Human Factors in Computing Systems (CHI) and The ACM SIGACCESS Conference on Computers and Accessibility (ASSETS), have released guidelines for creating accessible submissions.\footnote{See \href{http://chi2019.acm.org/authors/papers/guide-to-an-accessible-submission/}{http://chi2019.acm.org/authors/papers/guide-to-an-accessible-submission/} and \href{https://assets19.sigaccess.org/creating_accessible_pdfs.html}{https://assets19.sigaccess.org/creating\_accessible\_pdfs.html}} The ACM Digital Library\footnote{\href{https://dl.acm.org/}{https://dl.acm.org/}} provides some publications in HTML format, which is easier to make accessible than PDF~\cite{Graells2007EstudioDL}. \citet{Ribera2019PublishingAP} conducted a case study on DSAI 2016 (Software Development and Technologies for Enhancing Accessibility and Fighting Infoexclusion). The authors of DSAI were responsible for creating accessible proceedings and identified barriers to creating accessible proceedings, including lack of sufficient tooling and lack of awareness of accessibility. The authors recommended creating a new role in the organizing committee dedicated to accessible publishing. These policy changes have led to improvements in localized communities, but have not been widely adopted by all academic publishers and conference organizers.

Table~\ref{tab:prior_work} lists prior studies that have analyzed PDF accessibility of academic papers, and shows how our study compares. Prior work has primarily focused on papers published in Human-Computer Interaction and related fields, specific to certain publication venues, while our analysis tries to quantify paper accessibility more broadly.
\citet{Brady2015CreatingAP} quantified the accessibility of 1,811 papers from CHI 2010-2016, ASSETS 2014, and W4A, assessing the presence of document tags, headers, and language. They found that compliance improved over time as a response to conference organizers offering to make papers accessible as a service to any author upon request. \citet{Lazar2017MakingTF} conducted a study quantifying accessibility compliance at CHI from 2010 to 2016 as well as ASSETS 2015,
confirming the results of \citet{Brady2015CreatingAP}. They found that across 5 accessibility criteria, the rate of compliance was less than 30\% for CHI papers in each of the 7 years that were studied. The study also analyzed papers from ASSETS 2015, an ACM conference explicitly focused on accessibility, and found that those papers had significantly higher rates of compliance, with over 90\% of the papers being tagged for correct reading order and no criteria having less than 50\% compliance. This finding indicates that community buy-in is an important contributor to paper accessibility.
\citet{Nganji2015ThePD} conducted a study of 200 PDFs of papers published in four disability studies journals, finding that accessibility compliance was between 15-30\% for the four journals analyzed, with some publishers having higher adherence than others. To date, no large scale analysis of scientific PDF accessibility has been conducted outside of disability studies and HCI, due in part to the challenge of scaling such an analysis. We believe such an analysis is useful for establishing a baseline and characterizing routes for future improvement. Consequently, as part of this work, we conduct an analysis of scientific PDF accessibility across various fields of study, and report our findings relative to prior work.

\begin{table}[t!]
\small
    \centering
    \begin{tabularx}{\linewidth}{L{22mm}L{15mm}L{48mm}L{16mm}L{34mm}}
        \toprule
        \textbf{Prior work} & \textbf{PDFs analyzed} & \textbf{Venues} & \textbf{Year} & \textbf{Accessibility checker} \\
        \midrule
        \citet{Brady2015CreatingAP} & 1811 & CHI, ASSETS and W4A & 2011--2014 & PDFA Inspector \\ [0.5mm]
        \hline \\ [-2.5mm]
        \citet{Lazar2017MakingTF} & 465 + 32 & CHI and ASSETS & 2014--2015 & Adobe Acrobat Action Wizard \\ [0.5mm]
        \hline \\ [-2.5mm]
        \citet{Ribera2019PublishingAP} & 59 & DSAI & 2016 & Adobe PDF Accessibility Checker 2.0 \\ [0.5mm]
        \hline \\ [-2.5mm]
        \citet{Nganji2015ThePD} & 200 & \textit{Disability \& Society}, \textit{Journal of Developmental and Physical Disabilities}, \textit{Journal of Learning Disabilities}, and \textit{Research in Developmental Disabilities} & 2009--2013 & Adobe PDF Accessibility Checker 1.3 \\ [0.6mm]
        \hline \\ [-2.5mm]
        \textbf{\textit{Our analysis}} & \numpdfs & Venues across various fields of study & 2010--2019 & Adobe Acrobat Accessibility Plug-in Version 21.001.20145 \\
        \bottomrule
    \end{tabularx}
    \caption{Prior work has investigated PDF accessibility for papers published in specific venues such as CHI, ASSETS, W4A, DSAI, or various disability journals. Several of these works were conducted manually, and were limited to a small number of papers, while the more thorough analysis was conducted for CHI and ASSETS, two conference venues focused on accessibility and HCI. Our study expands on this prior work to investigate accessibility over \numpdfs PDFs sampled from across different fields of study.
    }
    \label{tab:prior_work}
\end{table}
\section{Analysis of academic PDF accessibility}
\label{sec:sos}

To capture and better characterize the scope and depth of the problems around academic PDF accessibility, we perform a broad meta-scientific analysis. We aim to measure the extent of the problem (e.g., what proportion of papers have accessible PDFs?), whether the state of PDF accessibility is improving over time (e.g., are papers published in 2019 more likely to be accessible than those published in 2010?), and whether the typesetting software used to create a paper is associated with the accessibility of its PDF (e.g., are papers created using Microsoft Word more or less accessible than papers created with other software?).

Prior studies on PDF accessibility have been limited to papers from specific publication venues such as CHI, ASSETS, W4A, DSAI, and journals in disability research. Notably, these venues are closer to the field of accessible computing, and are consequently more invested in accessibility.\footnote{See submission and accessibility guidelines for ASSETS (\href{https://assets19.sigaccess.org/creating_accessible_pdfs.html}{https://assets19.sigaccess.org/creating\_accessible\_pdfs.html}), CHI (\href{https://chi2021.acm.org/for-authors/presenting/papers/guide-to-an-accessible-submission}{https://chi2021.acm.org/ for-authors/presenting/papers/guide-to-an-accessible-submission}), W4A (\href{http://www.w4a.info/2021/submissions/technical-papers/}{http://www.w4a.info/2021/submissions/technical-papers/}) and DSAI (\href{http://dsai.ws/2020/submissions/}{http://dsai.ws/2020/submissions/}).}  We expand upon this work by investigating accessibility trends across various fields of study and publication venues. Our goal is to characterize the overall state of paper PDF accessibility and identify ongoing challenges to accessibility going forward.

\subsection{Data \& methods}\label{subsec:data-methods}

We sample PDFs from the Semantic Scholar literature corpus \citep{Ammar2018ConstructionOT} for analysis. We construct a dataset of papers by sampling PDFs published in the years of 2010--2019 stratified across the 19 top level fields of study defined by Microsoft Academic Graph \citep{msr:mag1, Shen2018AWS}. Examples of fields include Biology, Computer Science, Physics, Sociology, and others. This dataset allows us to investigate the overall state of PDF accessibility for academic papers, and to study the relationship between field of study and PDF accessibility. 

For each field of study, we sample papers from the top venues by total citation count, along with some documents without venue information, which include things like books and book chapters. The resulting papers come from 1058 unique publication venues; for each field of study, between 29 and 110 publication venues are represented, with Art on the minimum end, and Economics and Computer Science on the maximum end. Each field is represented by an average of 65 different publication venues. The vast majority of documents sampled into our dataset are published papers, rather than preprints or other non-peer-reviewed manuscripts. Publication venues represented in our sample are generally highly reputable journals, for example, \textit{The Lancet} or \textit{Neurology} for Medicine, \textit{The Astrophysical Journal} and \textit{Physical Review Letters} for Physics, or various IEEE publications for Computer Science and Engineering. In some cases, the mapping between publication venue and field of study can be unclear; for example, the publication venue \textit{Mathematical Problems in Engineering} is associated with Mathematics in our sample rather than Engineering. From an examination of the data, classifications seem reasonable and could be justified. We estimate that around 2.2\% of the sample are conference papers, 6.1\% are book chapters, reports, or lecture notes, less than 0.5\% are preprints, and the remaining majority are journal publications. We believe this is a reasonably representative sample of paper-like documents available to scholars and researchers.

We analyze the PDFs in our dataset using the Adobe Acrobat Pro DC PDF accessibility checker.\footnote{\href{https://www.adobe.com/accessibility/products/acrobat/using-acrobat-pro-accessibility-checker.html}{https://www.adobe.com/accessibility/products/acrobat/using-acrobat-pro-accessibility-checker.html}} Though this checker is proprietary and requires a paid license, it is the most comprehensive accessibility checker available and has been used in prior work on accessibility \citep{Lazar2017MakingTF, Ribera2019PublishingAP, Nganji2015ThePD}. Alternatively, non-proprietary PDF parsers such as PDFBox\footnote{\url{https://github.com/apache/pdfbox}} do not consistently extract accessibility criteria from sample PDFs, even when the criteria are met. We also prefer Adobe's checker to PDFA Inspector, used by \citet{Brady2015CreatingAP}, because PDFA Inspector only analyzes three criteria, whereas we are interested in other accessibility attributes as well, like the presence of alt-text.

For each PDF, the Adobe accessibility checker generates a report that includes whether or not the PDF passes or fails tests for certain accessibility features, such as the inclusion of figure alt-text or properly tagged headings for navigation. Because there is no API or standalone application for the Adobe accessibility checker, it can only be accessed through the user interface of a licensed version of Adobe Acrobat Pro. We develop an AppleScript program that enables us to automatically process papers through the Adobe checker. Our program requires a dedicated computer running MacOS and a licensed version of Adobe Acrobat Pro. It takes 10 seconds on average to download and process each PDF, which enables us to scale up our analysis to tens of thousands of papers. Accessibility reports from the checker are saved in HTML format for subsequent analysis.

Each report contains a total of 32 accessibility criteria, marked as ``Passed,'' ``Failed,'' or ``Needs manual check.''\footnote{Please see \href{https://helpx.adobe.com/acrobat/using/create-verify-pdf-accessibility.html}{https://helpx.adobe.com/acrobat/using/create-verify-pdf-accessibility.html} for a description of the accessibility report.}
Following \citet{Lazar2017MakingTF}, we analyze the following five criteria\footnote{For papers containing no tables and/or no figures, the Adobe checker can still return both pass or fail for the Table header and Alt-text criteria respectively. When objects in the PDF are \textit{not} tagged, the checker will fail these criteria even when the paper has no tables and/or no figures. When objects in the PDF \textit{are} tagged and the PDF is accessible, the checker will pass these criteria even when the paper has no tables or no figures.}:

\begin{itemize}
    \item Alt-text: Figures have alternate text.
    \item Table headers: Tables have headers.
    \item Tagged PDF: The document is tagged to specify the correct reading order.
    \item Default language: The document has a specified reading language.
    \item Tab order: The document is tagged with correct reading order, used for navigation with the \texttt{tab} key.
\end{itemize}

\noindent 
For our analysis, we also report \textit{Total Compliance}, which refers to the sum number of accessibility criteria met (e.g. if a paper has met 3 out of the 5 criteria we specify, then Total Compliance is 3). In some cases, we report the \textit{Normalized Total Compliance}, which is computed as the Total Compliance divided by 5, and can be interpreted as the proportion of the 5 criteria which are satisfied. We also report \textit{\xcompliance{5}}, a binary value of whether a paper has met all 5 criteria we specify (1 if all 5 criteria are met, 0 if any are not met), and the rate of \xcompliance{5} for papers in our dataset.

In addition to running the accessibility checker, we also extract metadata for each PDF, focusing on metadata related to the PDF creation process. PDF metadata are generated by the software used to create each file, and we analyze the associations between different PDF creation software and the accessibility of the resulting PDF document. Our hypothesis is that some classes of software (such as Microsoft Word) produce more accessible PDFs.

\subsection{Accuracy of our automated accessibility checker}
\label{sec:sos_chi}

\begin{table}[t!]
\begin{tabular}{lccc}
    \toprule
    \textbf{Criterion} & \textbf{CHI 2010\citep{Lazar2017MakingTF}} & \textbf{Ours-CHI 2010} & \textbf{Ours-All (\numpdfs)} \\ 
    \midrule
    Alt-text & 3.6\% & 4.0\% & 7.5\%  \\
    Table headers & 0.7\% & 1.0\% & 13.3\% \\
    Tagged PDF & 6.3\% & 7.4\% & 13.4\% \\
    Default language & 2.3\% & 3.0\% & 17.2\% \\
    Tab order & 0.3\% & 1.0\% & 9.3\% \\
    \midrule
    \xcompliance{5} & - & - & 2.4\% \\
    \bottomrule
\end{tabular}
\caption{We reproduce the analysis conducted by \citet{Lazar2017MakingTF} on PDFs of papers published in CHI, showing the percentage of papers that satisfy each of the five accessibility criteria. We find similar compliance rates, indicating that our automated accessibility checker pipeline is comparable to previous analysis methods. We also show the percentage of papers in our full dataset of \numpdfs PDFs that satisfy each criterion, along with the percent that satisfy \xcompliance{5}.
}
\label{tab:chi-results}
\end{table}

Previous work employed different versions of the Adobe Accessibility Checker to generate paper accessibility reports. To confirm the accuracy of our checker, as well as the automated script we create to perform the analysis, we run our checker on CHI 2010 papers to reproduce the results of \citet{Lazar2017MakingTF}. We identify CHI papers using DOIs reported by the ACM, and resolve these to PDFs in the Semantic Scholar corpus \citep{Ammar2018ConstructionOT}. We identify \numchi CHI papers in the corpus, and generate accessibility reports for these using our automated checker.

Our results shows similar rates of compliance compared to what was measured by \citet{Lazar2017MakingTF} (see Table~\ref{tab:chi-results} for results). This indicates that our automated accessibility checker produces comparable results to previous studies.

\subsection{Proportion of papers with accessible PDFs}
\label{sec:sos_fos}

\begin{figure}[tb!]
  \centering
    \includegraphics[width=0.55\linewidth]{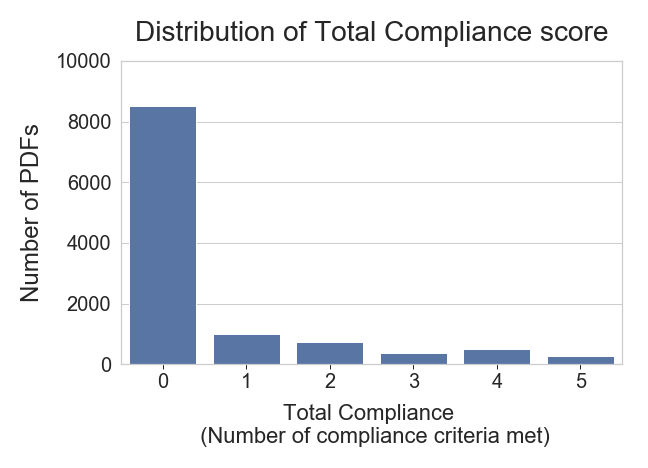}
  \caption{The distribution of numbers of PDFs in our dataset that meet our defined accessibility compliance criteria. A large majority (8519) of PDFs in our sample meet 0 out of 5 accessibility criteria. Of those meeting 1 criterion (Total Compliance = 1), the most commonly met criterion is Default Language (793 of 1010, 78.5\%). Of those meeting 4 criteria (Total Compliance = 4), the most common missing criterion is Alt-text (396 of 494, 80.2\%).
  } 
  \label{fig:fos-total-compliance}
  \Description{A histogram showing the distribution of total compliance score for our dataset. The majority of PDFs (8519 of 11397) in our sample meet 0 compliance criteria. Small numbers of PDFs meet some criteria, with lower numbers meeting more criteria.}
\end{figure}

\begin{figure}[t!]
  \centering
    \includegraphics[width=0.9\linewidth]{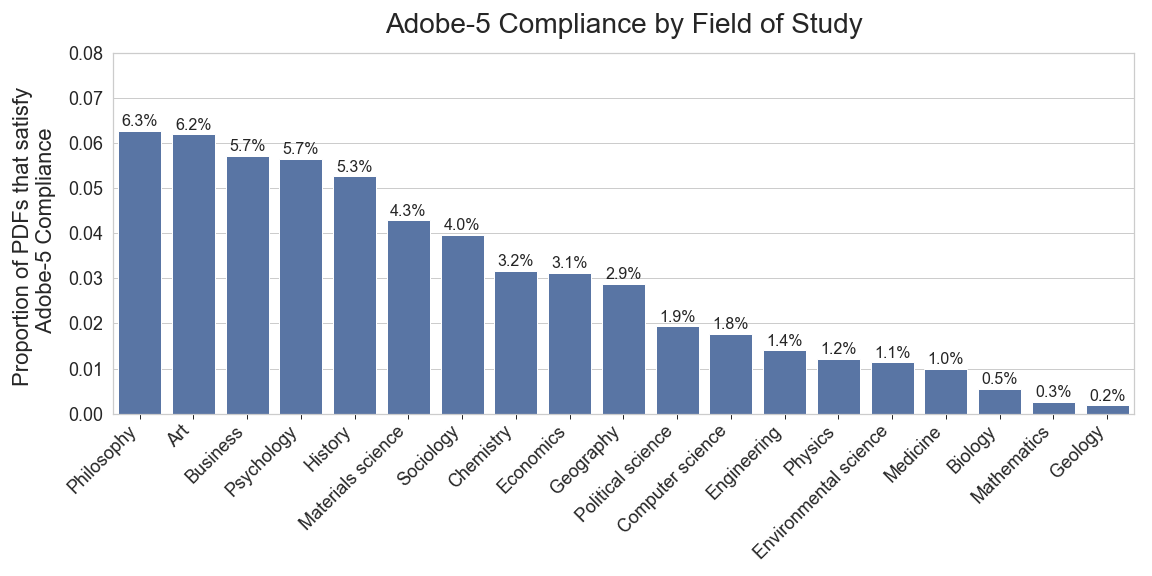}
  \caption{Percent of papers per field of study that meet all 5 accessibility criteria defined in \xcompliance{5}. Philosophy, Art, and Psychology have the highest rates of \xcompliance{5} satisfaction while Biology, Mathematics, and Geology have the lowest rates. 
  None of the fields had more than $6.5\%$ of PDFs satisfying \xcompliance{5}. 
  }
  \label{fig:fos-complete-compliance}
  \Description{A bar plot showing the proportion of PDFs in each field of study that satisfy Adobe-5 Compliance (meets all five accessibility criteria we define). Compliance percentage ranges from 6.3\% at the high end to 0.2\% at the low end. At the high end are fields such as philosophy, art, business, and psychology. At the low end are fields like biology, mathematics, and geology.}
\end{figure}

Around 1.6\% of PDFs we attempted to process failed in the Adobe checker (i.e., we could not generate an accessibility report). The accessibility checker most commonly fails because the PDF file is password protected, or the PDF file is corrupt. In both of these cases, the PDF is inaccessible to the user. We exclude these PDFs from subsequent analysis.

Accessibility compliance over all papers is low. Table~\ref{tab:chi-results} shows the percent of papers meeting each of the five criteria, as well as the Adobe-5 Compliance rate associated with this sample of papers. Figure~\ref{fig:fos-total-compliance} shows that the vast majority of papers do not meet any of the five accessibility criteria (8519 papers, 74.7\% do not meet any criteria) and very few (275 papers, 2.4\%) meet all five. Of those PDFs meeting 1 criterion, the most commonly met criterion is Default Language (793 of 1010, 78.5\%). Of those PDFs meeting 4 criteria, the most common \textit{missing} criterion is Alt-text (396 of 494, 80.2\%). In fact, only 854 PDFs (7.5\%) in the whole dataset have alt-text for figures. This is intuitive as Alt-text is the only criterion that \textit{always} requires author input to achieve, while the other four criteria can be derived from the document or automatically inferred, depending on the software used to generate the PDF.
    
As shown in Figure~\ref{fig:fos-complete-compliance}, all fields have an \xcompliance{5} of less than 7\%. The fields with the highest rates of compliance are Philosophy (6.3\%), Art (6.2\%), Business (5.7\%), Psychology (5.7\%), and History (5.3\%) while the fields with the lowest rates of compliance are Geology (0.2\%), Mathematics (0.3\%), and Biology (0.6\%). Fields associated with higher compliance tend to be closer to the humanities, and those with lower levels of compliance tend to be science and engineering fields. The prevalence of different document editing and typesetting software by field of study may explain some of these differences, and we explore these associations in Section~\ref{sec:sos_pdf_headers}.

\subsection{Trends in paper accessibility over time}

\begin{figure}[t!]
  \centering
    \includegraphics[width=0.6\linewidth]{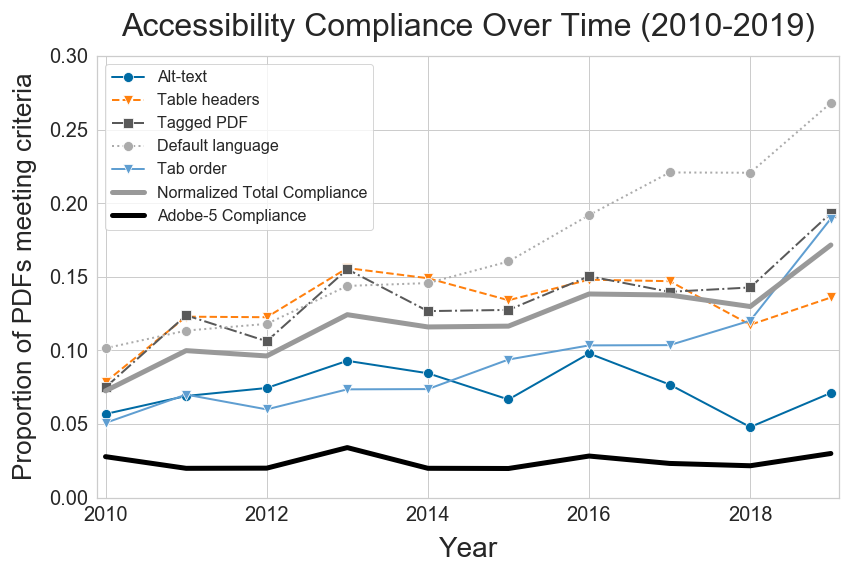}
  \caption{Accessibility compliance over time (2010-2019). The rate of \xcompliance{5} has remained relatively stable over the last decade, at around 2--3\%. Compliance along several criteria have improved over time, though the largest improvements have been in Default Language, the simplistic criteria to meet. Modest improvements are seen for Table headers, Tagged PDFs, and Tab order. The presence of alt-text has remained stable and lower, around 5--10\%. 
  }
  \label{fig:fos-over-time}
  \Description{A line plot shows changes in compliance rates over time. Adobe-5 Compliance has stayed consistently around 0.02-0.03 since 2010. The proportion of PDFs satisfying the Default language criteria has increased the most over time, from 0.10 in 2010 to 0.27 in 2019. Tagged PDF, Tab order, and Table headers also show improvements. The proportion of PDFs with alt-text has remained fairly consistent over time, between 0.05 and 0.1.}
\end{figure}

We show changes in compliance for all fields of study over time in Figure~\ref{fig:fos-over-time}. With the exception of Default Language, all accessibility criteria demonstrate slowly increasing or stable compliance rates over the past decade, with increases seen in Tagged PDFs and Tab order over time. Default language compliance is increasing most rapidly, from around 10\% compliance in 2010 to more than 25\% in 2019. This may be due to changes in PDF generation defaults in various typesetting software. Though this improvement is good, Default Language is the easiest of the five criteria to bring into compliance, and arguably the least valuable in terms of improving the accessible reading experience. The criterion with the lowest rate of compliance is Alt-text, which has remained stable between 5--10\% and has been lower in recent years. Since Alt-text is the only criterion of the five which always necessitates author intervention, we believe this is a sign that authors have not become more attuned to accessibility needs, and that at least some of the improvements we see over time can be attributed to typesetting software or publisher-level changes.

\begin{table}[t!]
\begin{tabular}{ll}
    \toprule
        \textbf{Typesetting Software} & \textbf{Count (\%)} \\ 
    \midrule
        Adobe InDesign       & 1591 (14.0\%) \\
        LaTeX                & 1431 (12.6\%) \\
        Arbortext APP        & 1374 (12.1\%) \\
        Microsoft Word       & 1318 (11.6\%) \\
        Printer              & 1021 (9.0\%) \\
    \midrule
        Other                & 4662 (40.9\%) \\
    \bottomrule
\end{tabular}
\caption{Count of papers per Typesetting Software. ``Other'' includes PDFs created with an additional 24 unique software programs, each with counts of less than 350, as well as those created with an unknown typesetting software.}
\label{tab:dist-of-headers}
\end{table}

\subsection{Association between typesetting software and paper accessibility}
\label{sec:sos_pdf_headers}

Typesetting software is extracted from PDF metadata and manually canonicalized. We extract values for three metadata fields: \texttt{xmp:CreatorTool}, \texttt{pdf:docinfo:creator\_tool}, and \texttt{producer}. All unique PDF creation tools associated with more than 20 PDFs in our dataset are reviewed and mapped to a canonical typesetting software. For example, the values (\texttt{latex}, \texttt{pdftex}, \texttt{tex live}, \texttt{tex}, \texttt{vtex pdf}, \texttt{xetex}) are mapped to the LaTeX cluster, while the values (\texttt{microsoft}, \texttt{for word}, \texttt{word}) and other variants are mapped to the Microsoft Word cluster. We realize that not all Microsoft Word versions, LaTeX distributions, or other versions of typesetting software within a cluster are equal, but this normalization allows us to generalize over these software clusters. For analysis, we compare the five most commonly observed typesetting software clusters in our dataset, grouping all others into a cluster called \texttt{Other}.

\begin{figure}[t!]
  \centering
    \includegraphics[width=0.5\linewidth]{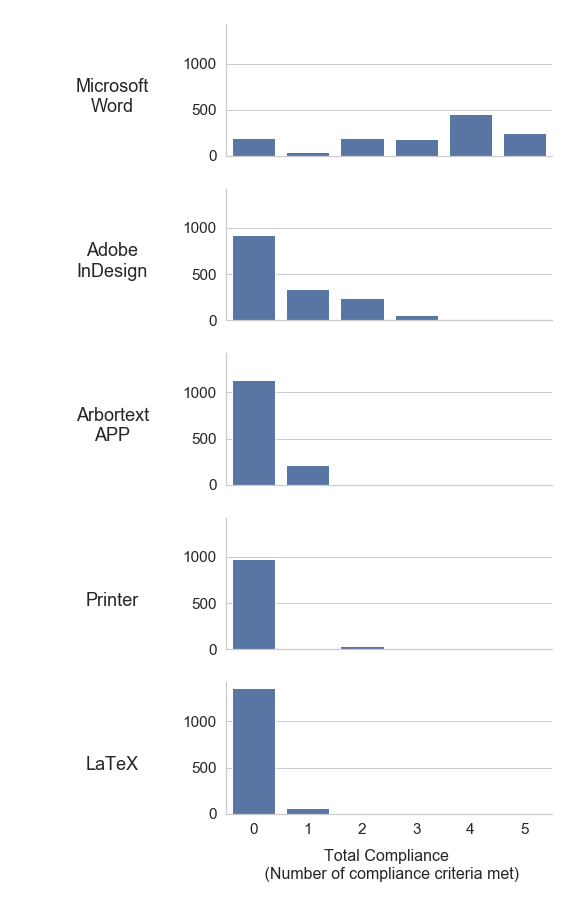}
  \caption{Histograms showing the distribution of Total Compliance scores for each of the top 5 typesetting software, ordered by decreasing mean Total Compliance. Microsoft Word stands out as producing PDFs with significantly higher Total Compliance than other typesetting software. Three of the top five PDF typesetting software clusters, Arbortext APP, Printer, and LaTeX, produce PDFs with low Total Compliance, with the majority of PDFs at 0 compliance. 
  }
  \Description{Five histograms show the distribution of Total Compliance scores for the five most common typesetting software clusters. These are sorted from most compliant to least compliant, in order: Microsoft Word, Adobe InDesign, Arbortext APP, Printer, and LaTeX. Microsoft Word produces many PDFs that satisfy 2 or more criteria, with a peak at Total Compliance = 4. Most PDFs produced by Adobe InDesign satisfy no accessibility criteria, but many satisfy 1 or 2. Arbortext APP, Printer, and LaTeX all produce inaccessible PDFs, with the vast majority of PDFs produced by these software satisfying no accessibility criteria.}
  \label{fig:fos-total-compliance-headers}
\end{figure}

We report the distribution of typesetting software in Table~\ref{tab:dist-of-headers}. The most popular PDF creators are Adobe InDesign, LaTeX, Arbortext APP, Microsoft Word, and Printer. ``Printer'' refers to PDFs generated by a printer driver (by selecting ``Print'' $\rightarrow$ ``Save as PDF'' in most operating systems); unfortunately, creating a PDF through printing provides no indicator of what software was used to typeset the document, and is generally associated with very low accessibility compliance. The ``Other'' category aggregates papers created by all other clusters of typesetting software; each of these clusters is associated with less than 350 PDFs, i.e., the falloff is steep after the Printer cluster. For the following analysis, we present a comparison between the five most common PDF creator clusters.

Figure~\ref{fig:fos-total-compliance-headers} shows histograms of the Total Compliance score for PDFs in the five most common typesetting software clusters. While the vast majority of papers do not meet any accessibility criteria, it is clear that Microsoft Word produces the most accessible PDFs, followed by Adobe InDesign. 
To determine the significance of this difference, we compute the ANOVA and Kruskal-Wallace \citep{Kruskal1952UseOR} statistics with the PDF typesetting software clusters as the sample groups and the Total Compliance as the measurements for the groups. We compute an ANOVA statistic of 2587.1 ($p$ < 0.001) and a Kruskal-Wallace $H$ statistic of 4422.0 ($p$ < 0.001). This indicates a significant difference in the distribution of Total Compliance scores between the five most common PDF typesetting software.

\begin{figure}[t!]
  \centering
    \includegraphics[width=0.64\linewidth]{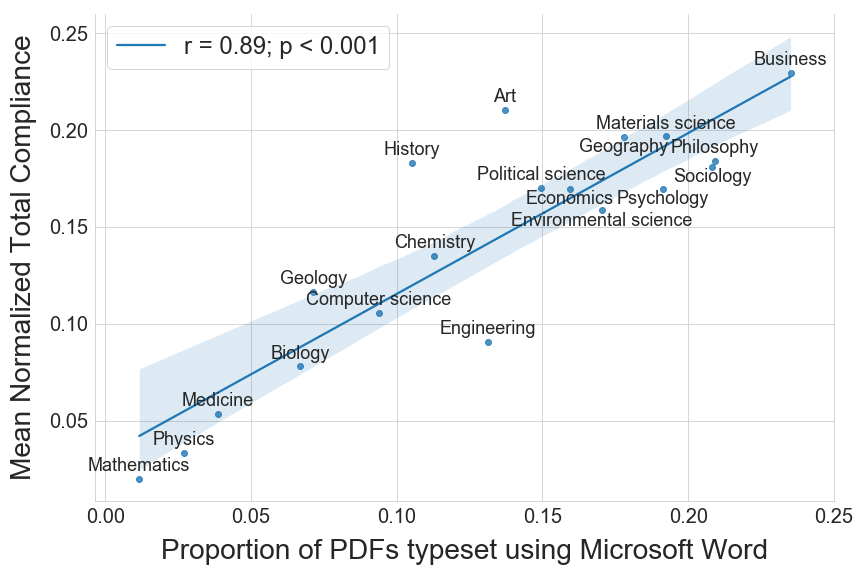}
  \caption{There is a strong correlation ($r = 0.89$, $p < 0.001$, 95\% CI shown) between the proportion of PDFs typeset using Microsoft Word and the mean normalized Total Compliance of papers by field of study. Fields such as Business, Philosophy, Sociology, Materials science, and Psychology use Microsoft Word around or over 20\% of the time, and have correspondingly higher mean accessibility compliance. On the other end of the spectrum are fields like Mathematics, Physics, and Medicine, where Microsoft Word is rarely used, and which have very low levels of mean compliance.}
  \label{fig:prop_word_by_fos}
  \Description{A scatter plot shows a positive correlation between the proportion of PDFs typeset using Microsoft Word and the mean normalized total compliance score by field of study. Fields that typeset more using Word have higher Total Compliance scores. The correlation coefficient r is 0.89, and p is less than 0.001. Fields that use Word more and have higher compliance rates include Business, Materials science, Geography, Philosophy, and Sociology. Fields that use Word very little and have low compliance rates include Mathematics, Physics, and Medicine.}
\end{figure}

In Figure~\ref{fig:prop_word_by_fos}, we observe again that usage of Microsoft Word is highly correlated with accessibility compliance. Here, we plot the proportion of Microsoft Word usage per field of study and the corresponding mean normalized Total Compliance rates for those fields. Higher rates of Microsoft Word usage are statistically correlated with higher mean normalized Total Compliance ($r = 0.89$, $p < 0.001$). 

\begin{figure}[t!]
  \centering
    \includegraphics[width=0.6\linewidth]{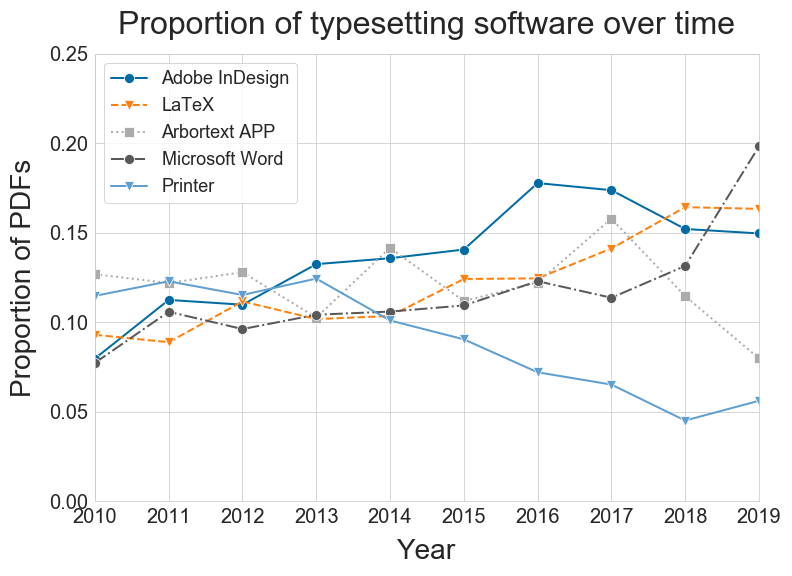}
  \caption{The proportion of PDFs typeset by the five most common typesetting software over time. Software such as Adobe InDesign, LaTeX, and Microsoft Word are increasing in popularity over time.}
  \label{fig:software_over_time}
  \Description{A line plot shows the proportion of different typesetting software in our sample between 2010-2019. Microsoft Word, Adobe InDesign, and LaTeX have all increased in proportional usage (all starting at a proportion of 0.07-0.08 in 2010 to 0.15 or above in 2019. The usage of printer drivers to create PDFs has declined from a proportion of 0.12 in 2010 to around 0.05 in 2019. Arbortext APP also shows a modest decline in recent years to below a proportion 0.10.}
\end{figure}

In Figure~\ref{fig:software_over_time}, we show the proportion of usage of each of the five typesetting software over time. In recent years, Adobe InDesign, LaTeX, and Microsoft Word usage are proportionally increasing, while the proportion of Printer-created PDFs is declining. The increase in Adobe InDesign and Microsoft Word have likely driven the increase in rates of Total Compliance over time, since these typesetting software are the most associated with higher accessibility compliance.

\subsection{Summary of analyses}
\label{sec:sos_summary}

Overall, accessibility compliance over the past decade and across all fields of study have slowly improved. Full compliance based on \xcompliance{5}, however, has remained around 2.4\% on average and does not show trends towards improving. Improvements in several compliance criteria are observed, with Default Language being the most improved, nearing 30\% coverage in 2019. However, Default Language is the easiest criteria to meet, and arguably produces the least amount of accessibility improvement in user experience. Criteria such as Tagged PDFs, Tab order, and Headers show modest improvements over time, though only between 10--15\% of papers in our sample meet any one of these individual criteria. Alt-text compliance is the lowest of our measured criteria, and as the only criterion of the five requiring author intervention in all cases, the lack of alt-text may be indicative of the general lack of author awareness and contribution to accessibility efforts for scientific papers. 

Based on our analysis, typesetting software plays a large role in document accessibility. Of the most common PDF creator software, Microsoft Word appears to produce the most accessibility-compliant PDFs, while LaTeX produces PDFs with the lowest compliance. Microsoft has recently made investments in the accessibility of their Office 365 Suite.\footnote{\href{https://www.microsoft.com/en-us/accessibility/microsoft-365}{https://www.microsoft.com/en-us/accessibility/microsoft-365}} It is clear that software can help increase accessibility compliance by prioritizing accessibility concerns during document creation, and we encourage other developers of typesetting and publishing software to priotize accessibility concerns in their development process. 

Improvements in accessibility compliance have stalled over the past decade, likely because accessibility concerns are considered marginal, and are outside of the awareness of most publishing authors and researchers. Significant changes in the authorial and publication processes are needed to change this status quo, and to increase the accessibility of scientific papers for BLV users going forward. Though we believe and encourage change in the academic paper authorial and publication process in relation to accessibility, the likelihood of rapid improvement is low and these changes will not impact the many millions of academic PDFs that have already been published. Therefore, we introduce a technological solution that may mitigate some of the accessibility challenges of existing paper PDFs, and aim to understand how this solution and others like it could serve the immediate needs of the BLV research community.

\section{Converting PDF to HTML: The \scially Pipeline}
\label{sec:pdf2html}

To address the broad accessibility challenges described in Section \ref{sec:sos}, we propose and prototype a system for extracting semantic content from paper PDFs and re-rendering this content as accessible HTML. HTML is widely accepted as a more accessible document format than PDFs. In the 2019 Access SIGCHI Report, the authors discuss the reasoning behind switching CHI publications to a new HTML5 proceedings format to improve accessibility \citep{Mankoff2019SIGCHI}. 
By rendering the content of paper PDFs as HTML, and introducing proper reading order and accessibility features such as section headings, links, and figure tags, we can offset many of the issues of reading from an inaccessible PDF. Our PDF to HTML rendering system is named \scially after the community-adopted numeronym for digital accessibility.\footnote{\href{https://www.a11yproject.com/}{https://www.a11yproject.com/}} 

Figure~\ref{fig:pipeline} provides a schematic for the approach. \scially leverages the two open source PDF processing projects S2ORC \citep{lo-wang-2020-s2orc} and DeepFigures \citep{Siegel2018ExtractingSF}, the Semantic Scholar API,\footnote{\href{https://api.semanticscholar.org/}{https://api.semanticscholar.org/}} and a custom Flask application for rendering the extracted content of the PDF as HTML. The S2ORC project \citep{lo-wang-2020-s2orc} integrates the Grobid machine learning library \citep{Lopez2015GROBIDI} and a custom XML to JSON parser\footnote{Available at \href{https://github.com/allenai/s2orc-doc2json}{https://github.com/allenai/s2orc-doc2json}} to produce a structured representation of paper text. We use a version of the S2ORC pipeline that is based on Grobid v0.6.0. The resulting JSON representation includes metadata fields like title, authors, and affiliations, and paper content fields such as abstract, section headers, body text organized into paragraphs, bibliography entries, and figure and table objects (though not the figure images themselves). The output also contains links between inline citations and figure/table references respectively to bibliography entries and figure/table objects. DeepFigures \citep{Siegel2018ExtractingSF}, on the other hand, leverages a computer vision model to extract images of figures and tables as well as their corresponding captions from the source PDF. 

The outputs of S2ORC and DeepFigures are stitched together to form the HTML render as in Figure~\ref{fig:pipeline}. We place header tags (\texttt{<h1>...</h1>}, \texttt{<h2>...</h2>}) around the title, authors, abstract, section headings, and reference heading. Paragraphs of body text are enclosed in \texttt{<p>...</p>} tags in order within their appropriate sections. Bibliography entries are provided in an unordered list under the reference heading. Figures and tables are enclosed in \texttt{<figure>...} \texttt{</figure>} tags and placement is inferred based on mentions in the text. A figure or table is placed immediately after the paragraph in which its handle is first mentioned (e.g. ``In \underline{Fig. 1}, we show...'' is the first mention of Figure 1 and the figure is placed directly after the paragraph with this mention). Figure and table captions are attached to their corresponding image objects, so that correspondences between the caption text and image are made explicit (in PDFs, this is usually not the case). Any figures or tables which are not mentioned in order in the text are placed in order nonetheless; in other words, if paragraph 1 mentions Figure 1 and paragraph 2 mentions Figure 3, both Figure 1 and 2 will be placed directly following paragraph 1 and Figure 3 following paragraph 2. This ensures that the layout for the HTML render closely approximates the intended reading order. We justify this decision based on user feedback from our pilot study, which is discussed in Section~\ref{sec:user_study}. 

\begin{figure}
    \centering
    \includegraphics[width=0.4\textwidth]{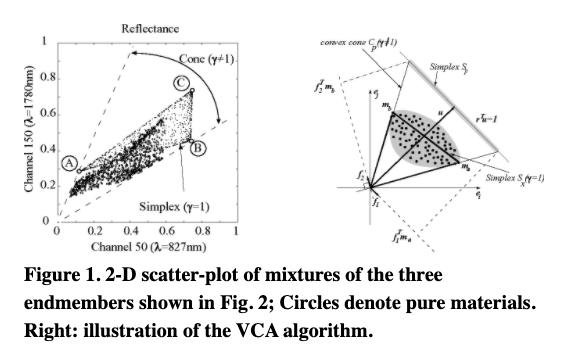}\includegraphics[width=0.4\textwidth]{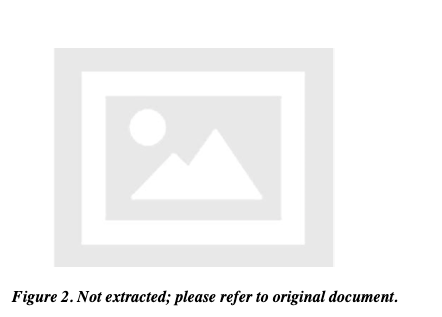}
    \includegraphics[width=0.6\textwidth]{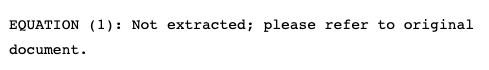}
    \caption{A successfully extracted figure from \citet{Nascimento2005VertexCA} is shown with its corresponding figure caption (\textit{top left}). When figures are not extracted and inferred to exist (handle mentioned in text or number between two extracted figures), a placeholder image is shown along with a message referencing the failed extraction (\textit{top right}). Similarly, when an equation is detected to be present in the PDF and not extracted, we insert text signaling the failed extraction and refer the user to the source document (\textit{bottom}).}
    \label{fig:figure_equations}
    \Description{Three subfigures show figure and equation related features in SciA11y. First subfigure shows extracted images from Nascimento and Bioucas-Dias and the associated caption of the figure, as displayed in an HTML <figure> block. Second subfigure shows a placeholder image for a figure that was not successfully extracted by the system. Third subfigure shows the text used as a placeholder for equations that are not extracted. Both the not extracted figure and not extracted equation are followed by the text: "Not extracted; please refer to original document."}
\end{figure}

In some cases, we are able to successfully process a PDF through S2ORC to extract textual content but DeepFigures either fails to process the PDF or fails to extract some or all figures from the PDF. To mitigate the cognitive dissonance around figure or table mentions without corresponding figure or table objects, we insert placeholder objects into the HTML render as in Figure~\ref{fig:figure_equations}. For example, if ``Figure 2'' is mentioned in the text but is not successfully extracted by DeepFigures, we would insert a placeholder image for the figure based on the logic described in the previous paragraph along with the text ``Figure 2. Not extracted; please refer to original document.'' Similarly, mathematical equations that we cannot currently extract are acknowledged with the same placeholder text.

We add links between inline citations and the corresponding reference entry where possible. We insert links at each inline citation in the body text that link to the corresponding bibliography entry. Following each bibliography entry, we provide links back to the first mention of that entry in each section of the paper in which it was mentioned. For example, if bibliography entry \texttt{[1]} is cited in the ``II. Related Works'' section and the ``III. Methods'' section, we provide two links following the entry in the bibliography to the corresponding citation locations in sections II and III, as in:

\begin{alltt}
    [1] Last name et al. Paper title. Venue. DOI.
        \underline{Link to return to Section II}, \underline{Link to return to Section III}
\end{alltt}

\noindent This allows users to navigate back to their reading location in the document after clicking through to a bibliography entry. A user may otherwise hesitate to resolve a link, because it may result in losing their place and train of thought. Finally, we introduce a table of contents near the beginning of the HTML render to facilitate better understanding of overall document structure. The table of contents includes all section titles, linked to the corresponding sections, as well as figures and tables nested under their respective section headers. The table of contents provides a rapid overview of the structure of the document, and facilitates rapid navigation to the reader's desired sections.

In the current iteration of the HTML render, we do not display author affiliations, footnotes, or mathematical equations due to the difficulty of extracting these pieces of information from the PDF. Though some of the elements are extracted in S2ORC, the overall quality of the extractions for these elements is lower, and is currently insufficient for surfacing in the prototype (see Section~\ref{sec:evaluation} for details). Future work includes investigating the possibility of extracting and exposing these elements, either by improving current models or training new models targeted towards the extraction of specific paper elements.

We leverage the feedback we received during our pilot studies (see Section~\ref{sec:user_study}) to make improvements prior to the main user study. We denote the versions of the prototype as v0.1 (initial version; version seen by P1), v0.2 (version seen by P2), and v0.3 (version seen by all other participants in the main user study). Features implemented in v0.1 include the primary components of the HTML render such as title, authors, abstract, body text with section headers, figures and tables, references, and links between inline citations and references. In v0.1, figures and tables were placed in a separate section following the main body of the paper. Following P1, for version v0.2, we implemented the table of contents, inserted placeholders for objects that we could not extract, and began inserting figures and tables into the body text adjacent to their first mentions. This last change was made in response to P1's feedback that navigating away to figures caused him to lose his reading location. Following P2, for version v0.3, we implemented only minor changes. P2 signaled during his session that URLs in the bibliography were not being correctly extracted, so we patched the data to correctly extract and display URLs in bibliography entries.

Based on our evaluation of the quality of these HTML renders (Section~\ref{sec:evaluation}) and user feedback and response (Section~\ref{sec:user_study}), we believe our approach can dramatically increase the screen reader navigability and accessibility of scientific papers across all disciplines by providing an alternate and more accessible HTML version of these papers. Properly tagged section headings allow for quick navigation and skimming of a paper, links between inline citations and bibliography entries allow users to browse to cited papers without losing their place, and figure tags for figure and table objects allow for direct navigation to these in-paper objects. We now discuss the quality of our PDF extractions (Section~\ref{sec:evaluation}) and user response to the prototype (Section~\ref{sec:user_study}) in detail.
\section{HTML render quality evaluation}
\label{sec:evaluation}

Extracting semantic content from PDF is an imperfect process. Though re-rendering a PDF as HTML can increase a document's accessibility, the process relies on machine learning models that can make mistakes when extracting information. As we glean from user studies, BLV users may have some tolerance for error, but there is an inherent trade-off between errors and perceived trust in the system. We conduct a study to estimate the (1) faithfulness of the HTML renders to the source PDFs, and (2) overall readability of the resulting HTML renders. We define \textit{faithfulness} as how accurately the HTML render represents different facets of the PDF document, such as displaying the correct title, section headers, and figure captions. These facets are measured as the number of errors that are made in rendering, e.g., mistakenly parsing one figure caption into the body text is counted as one error towards that facet. \textit{Readability}, on the other hand, is an ordinal variable meant to capture the overall usability of the parse. Documents are given one of three grades, those with no major problems, some problems, and many problems impacting readability.

To evaluate readability and faithfulness, we first perform open coding on a small sample of document PDFs and corresponding \scially HTML renders. The purpose of this exercise is to identify facets of extraction that impact the ability to read a paper. A rubric is then designed based on these identified facets. The process taken to design the evaluation rubric, the rubric's content, and annotation instructions are detailed in Section~\ref{sec:eval_rubric}. We then annotate a sample of \numeval papers across different fields of study using this rubric. For each category of errors identified during open coding, we compute the overall error rates seen in our sample. We also present the overall assessed readability, reported in aggregate over our sample and by fields of study. Results of this evaluation are presented in Section~\ref{sec:eval_res}.

\subsection{Open coding of document facets}
\label{sec:eval_coding}

One author performed open coding on a sample of papers, comparing the PDF and \scially HTML renders to identify inconsistencies and facets that impact the faithfulness of document representation. Papers are sampled from the Semantic Scholar API\footnote{\href{https://api.semanticscholar.org/}{https://api.semanticscholar.org/}} using various search terms, and selecting the top 3 results for each search term for which a PDF and S2ORC parse are available. Search terms were selected to achieve coverage over different domains, and the top papers are sampled to select for relevant publications. The author stopped sampling papers upon reaching saturation, resulting in 8 search terms and 24 papers. The search terms used were: \texttt{human computer interaction}, \texttt{epilepsy}, \texttt{quasars}, \texttt{language model}, \texttt{influenza epidemiology}, \texttt{anabolic steroids}, \texttt{social networks}, and \texttt{arctic snow cover}.

For each paper, the author evaluated the PDF and HTML render side-by-side, scanning through the document to identify points of difference between the two document representations. Specifically, the author looked for any text in the PDF that is not shown in the HTML, any text from the PDF that is mixed into the main text of the HTML (e.g. figure captions, headers, or footnotes that should be separate from the main text but are mixed in, interrupting the reading flow), and other parsing mistakes (e.g. errors with math, missing lists and tables etc). These observations are detailed qualitatively, and each facet is assessed for its faithfulness to the original PDF document as well as its overall impact on readability.

\subsection{Evaluation rubric}
\label{sec:eval_rubric}

Observations from open coding are coalesced into an evaluation rubric and form for grading the quality and faithfulness of the HTML render.
The evaluation form attempts to capture errors in PDF extraction that affect each of the primary semantic categories identified for proper reading. These semantic categories and common extraction errors are given in Table~\ref{tab:eval_cats}.

Questions in the form are designed to capture each type of faithfulness error, while allowing annotators to qualify their responses. We also include a question to capture the overall readability of the HTML render. Instructions for completing the annotation form are provided in Appendix~\ref{app:eval_instructions}; the final version of the form is replicated in Appendix~\ref{app:eval_instructions}; and the rubric for overall readability evaluation is given in Appendix~\ref{app:quality_rubric}. 

Three authors iterated twice on the content of the evaluation form, until they came to a consensus that all evaluation categories were adequately addressed using a minimum set of questions. Two authors then participated in pilot annotations, where each person independently annotated the same set of five papers sampled from the set labeled by the third author during open coding. Answers to all numeric questions were within $\pm 1$ for these five papers when comparing the two authors' annotations. All three authors discussed discrepancies in overall readability score, iterating on the rubric defined in Appendix~\ref{app:quality_rubric} and coming to a consensus. The finalized form and rubric are used for evaluation.

\begin{table}[t!]
    \small
    \centering
    \begin{tabularx}{\linewidth}{L{20mm}L{65mm}X}
        \toprule
        \textbf{Category} & \textbf{Description} & \textbf{Common errors} \\
        \midrule 
        \textsc{title} & The title and subtitle of the paper & Missing words \newline Extra words \\ 
        \midrule
        \textsc{authors} & A list of authors who wrote the paper; this includes affiliation, though we do not explicitly evaluate affiliation in this study & Missing authors \newline Extra authors \newline Misspellings \\
        \midrule
        \textsc{abstract} & The abstract of the paper & Some text not extracted \newline Other text incorrectly extracted as abstract \\
        \midrule
        \textsc{section headings} & The text of section headings & Some headings not extracted (part of body text) \newline Other text incorrectly extracted as headings \\
        \midrule
        \textsc{body text} & The main text of the paper, organized by paragraph under each section heading & Some paragraphs not extracted (missing) \newline Some text not extracted \newline Other text incorrectly extracted as body text \\
        \midrule
        \textsc{figures} & Images, captions, and alt-text of each figure & Figure not extracted \newline Caption text not extracted (part of body text) \newline Other text incorrectly extracted as caption text \\
        \midrule
        \textsc{tables} & Caption/title and content of each table & Table not extracted (not part of body text) \newline Table not extracted (part of body text) \newline Caption text not extracted (part of body text) \newline Other text incorrectly extracted as caption text  \\
        \midrule
        \textsc{equations} & Mathematical formulas, represented in TeX or Math ML; note: our current pipeline does not extract math & Some equations not extracted \newline Some equations incorrectly extracted \\
        \midrule
        \textsc{bibliography} & Bibliography entries in the reference section & Some bibliography entries not extracted \newline Some bibliography entries incorrectly extracted \newline Other text incorrectly extracted as bibliography \\ 
        \midrule
        \textsc{inline citations} & Inline citations from the body text to papers in the bibliography section & Some inline citations not detected \newline Some inline citations incorrectly linked \\
        \midrule
        \textsc{headers, footers \& footnotes} & Page headers and footers, footnotes, endnotes, and other text that is not a part of the main body of the document & Some headers and footers incorrectly extracted into body text \\
        \bottomrule
    \end{tabularx}
    \caption{Categories of paper objects identified for evaluation along with the common errors seen for each category.}
    \label{tab:eval_cats}
\end{table}

Of the categories and errors described in Table~\ref{tab:eval_cats}, our current pipeline does not extract table content and equations. Tables are extracted as images by DeepFigures \citep{Siegel2018ExtractingSF}, which do not contain table semantic information. Regarding equations, we distinguish between inline equations (math written in the body text) and display equations (independent line items that can usually be referenced by number); for this work, we evaluated a small sample of papers for successful extraction of display equations. Though some display equations are recognized, the quality of equation extraction is low, usually resulting in missing tokens or improper math formatting. Therefore, we decided to replace display equations in the prototype with the equation placeholder shown in Figure~\ref{fig:figure_equations}. Since problems with mathematical formulae are among those most mentioned by users in our study, equation extraction is among our most urgent future goals, and we discuss some options in Section~\ref{sec:future_work}.

\subsection{Evaluation results}
\label{sec:eval_res}

We start with the dataset of \numpdfs papers we analyze in Section~\ref{sec:sos},
and subsample 535 documents stratified by field of study. Two expert annotators with undergraduate science training code papers from this sample, with an aim of annotating around 20 papers per field of study. Though we achieve the target number for most fields, we missed this target for some fields closer to the humanities because more of these documents are difficult to manually annotate within our time and resource constraints.
For example, documents are deemed unsuitable for annotation if they are not papers (i.e., they are books, posters, abstracts, etc), if they are too long, or if they are not in English. In these cases, the annotators can skip the document. Detailed guidance on suitability is provided in the annotation instructions (see Appendix~\ref{app:eval_instructions}). 

The two annotators annotated \numeval unique papers and skipped \numevalskipped. The resulting annotated sample consists of papers from 195 unique publication venues. Each paper takes 5--10 minutes to grade. Documents are skipped primarily due to language (paper not in English), length, or the document is not a paper. Inter-annotator agreement is computed over a sample of 20 papers over each of the evaluated facets. We report Cohen's Kappa for categorical questions such as those on the extraction of title, authors, abstract, and bibliography. For numerical questions such as counting the occurrence of extraction errors related to figures, tables, section headings, and body paragraphs etc, we report the intraclass correlation coefficient (ICC) as well as the average difference of values between the two annotators. See Table~\ref{tab:eval_iaa} for these results.
Agreement was high for most element-level annotator questions. Annotators had the highest levels of disagreement on the evaluation of header/footer/footnote errors, section heading errors, and body paragraph errors, likely due to these being text-based and the most numerous; though the average differences reported between annotators on these questions are only between 1-2. Likewise, agreement on overall readability score is modest, at 0.55; we note, however, that neither annotator labeled any paper as having no major readability problems when the other annotator labeled it as having lots of readability problems.

\begin{table}[tb!]
    \centering
    \begin{tabularx}{0.94\linewidth}{L{40mm}L{15mm}L{15mm}L{15mm}L{15mm}L{25mm}}
        \toprule
        Evaluation criteria & Number of classes & Agreement & Cohen's Kappa & ICC & Mean Difference ($\pm$ SD) \\
        \midrule
        Title & 3 & 0.87 & 0.33 & - & - \\
        Authors & 3 & 1.00 & 1.00 & - & - \\
        Abstract & 3 & 0.95 & 0.64 & - & - \\
        \midrule
        Number of figures & - & 1.00 & - & 1.00 & 0.00 $\pm$ 0.00 \\
        Figure extraction errors & - & 0.89 & - & 1.00 & 0.11 $\pm$ 0.31 \\
        Figure caption errors & - & 0.89 & - & 1.00 & 0.11 $\pm$ 0.31 \\
        Number of tables & - & 0.92 & - & 0.98 & 0.12 $\pm$ 0.43 \\
        Table extraction errors & - & 0.89 & - & 0.98 & 0.17 $\pm$ 0.50 \\
        Table caption errors & - & 0.78 & - & 0.94 & 0.33 $\pm$ 0.67 \\
        \midrule
        Header/footer/footnote errors & - & 0.40 & - & 0.60 & 1.88 $\pm$ 2.12 \\
        Section heading errors & - & 0.71 & - & 0.79 & 0.71 $\pm$ 1.70 \\
        Body paragraph errors & - & 0.46 & - & 0.66 & 1.50 $\pm$ 2.22 \\
        \midrule
        Bibliography extraction & 4 & 0.94 & 0.82 & - & - \\
        Inline citation linking & 4 & 0.80 & 0.11 & - & - \\
        \midrule
        Overall score & 3 & 0.55 & 0.07 & - & - \\
        \bottomrule
    \end{tabularx}
    \caption{Inter-rater agreement for evaluation. For categorical questions, such as title, author, abstract, bibliography, inline citation, and overall score, we report the number of classes available for annotation, along with annotator agreement and Cohen's Kappa. For numerical questions, such as the number of each type of extraction error, we report agreement, the intraclass correlation coefficient (ICC), and the average difference and standard deviation of the values between the two annotators.}
    \label{tab:eval_iaa}
\end{table}

\begin{figure}[th!]
    \centering
    \includegraphics[width=0.9\linewidth]{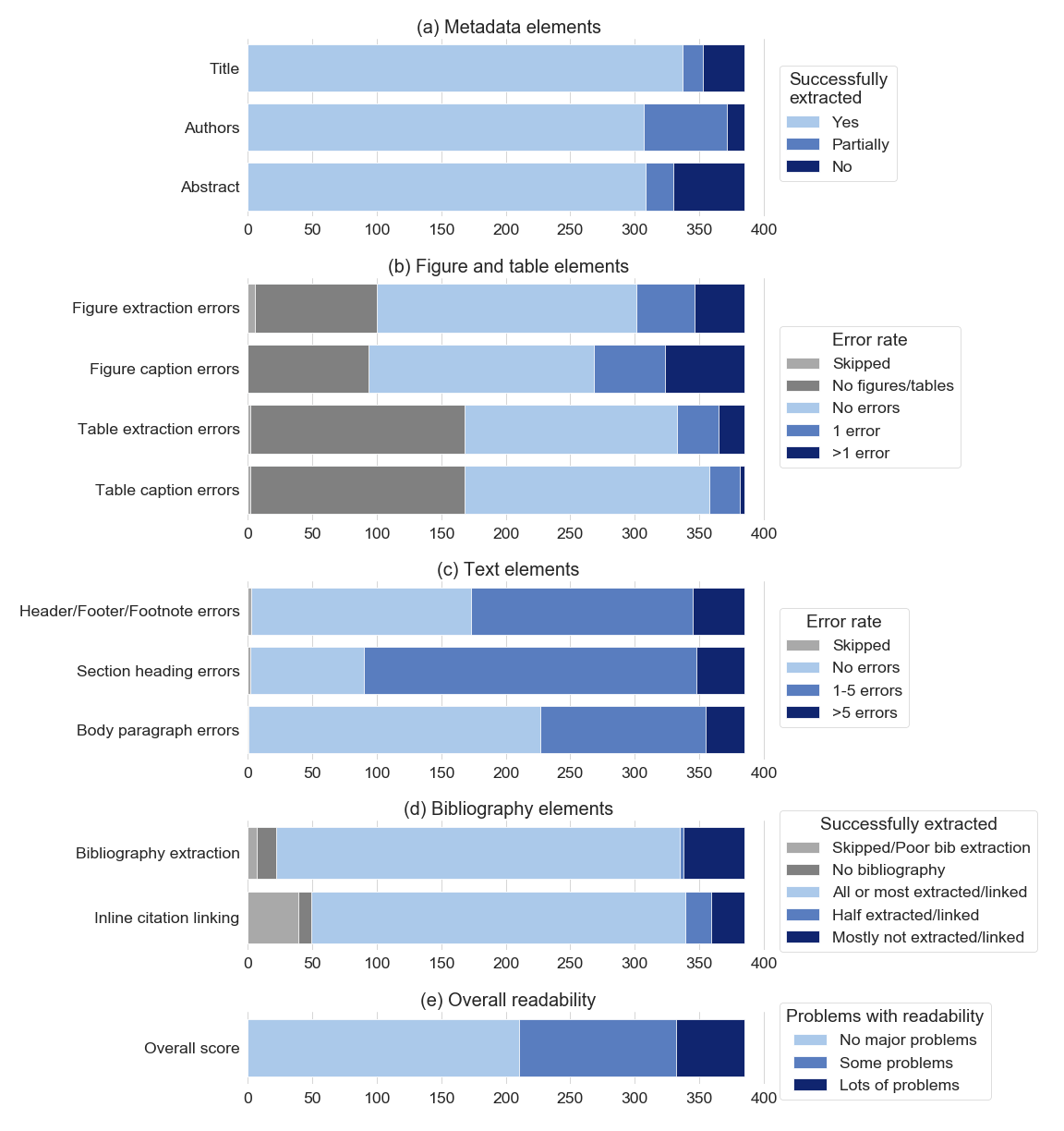}
    \caption{Evaluation results for various document components. Corresponding numbers are provided in Table~\ref{tab:eval_raw_by_element} in Appendix~\ref{app:eval_raw_results}.}
    \label{fig:eval_results}
    \Description{Percent stacked bar plots showing evaluation results. Metadata elements (title, authors, abstract) are extracted correctly the vast majority of the time. Around a quarter of papers don't have figures and two fifths don't have tables. Of those with figures, the majority are extracted correctly, though extraction errors for both images and captions are not infrequent. Of those with tables, table extraction errors and table caption extraction errors are infrequent. Text element errors (header/footer/footnote, section heading, and body paragraph) are the most frequent. For header/footer/footnotes, a bit more than half of all papers have one or more errors. The majority of papers have more than 1 section heading extraction error. A bit less than half of papers have body paragraph errors that affect 1 or more paragraphs. Bibliography extraction and inline citation linking are both good, with the vast majority of papers having all or most entries extracted and linked. For overall readability, more than half of evaluated papers have no major problems, around a third have some problems, and the remaining lots of problems.}
\end{figure}

\begin{figure}[th!]
    \centering
    \includegraphics[width=0.8\linewidth]{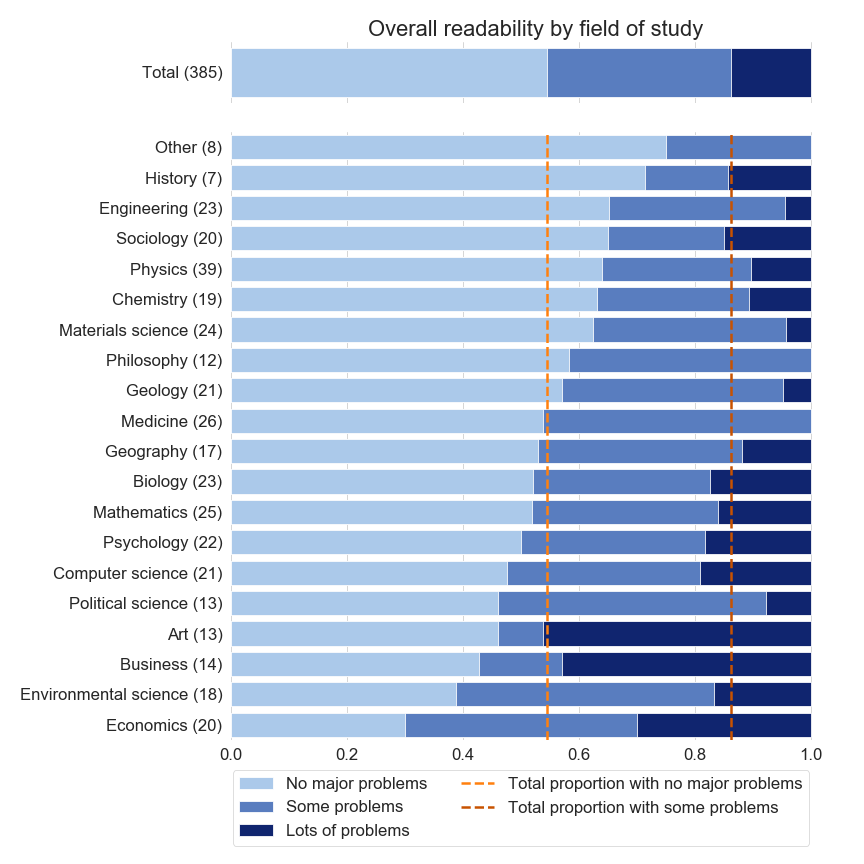}
    \caption{Overall readability results as proportion of total split by field of study, sorted by the percentage of papers with no major problems. The number of documents analyzed in each field is given, ranging from N=7 (History) to N=39 (Physics). The fields of study with the worst parse quality (Economics, Environmental science, Business, Art, and Political science) tend to be closer to the humanities, and may be due to the under-representation of papers from these fields in the data used to train the PDF parsers we use in our extraction pipeline. Corresponding numbers are provided in Table~\ref{tab:eval_raw_by_fos} in Appendix~\ref{app:eval_raw_results}.}
    \label{fig:eval_fos}
    \Description{Percent stacked bar plots show evaluation results for overall readability split by field of study. The bars are sorted on the field of study from those with the largest proportion of papers having no major problems to those with the least proportion having no major problems. Fields with the highest overall readability include history, engineering, sociology, physics, and chemistry. Fields with the lowest overall readability include political science, art, business, environmental science, and economics. No field stands out as being obviously different from the overall sample, though art, business and economics have the largest proportion of papers that were classified as having lots of readability problems.}
\end{figure}

All results and statistics are reported on the set of \numeval annotated papers. Figure~\ref{fig:eval_results} shows the breakdown of each type of error and the frequency at which it occurs. Metadata elements like title, authors, and abstract are successfully extracted the majority of the time. For figure and table elements, approximately 25\% of papers in our evaluation sample do not include figures, and around 45\% do not have tables. Of those that have figures, the majority (201, 69.1\% of 291) do not have extraction or parsing errors; around half of documents with errors have errors that only relate to one figure. Similarly, the majority of tables and table captions are correctly identified as tables and table captions, and are not incorrectly mixed into the body text. We note that the lack of an error here does not indicate that the table is extracted correctly in an accessible manner, just that it is not incorrectly parsed as body text.

Unsurprisingly, errors in text element parsing are the most prevalent, especially for headers/footers/footnotes and section headings. The most common type of header/footer/footnote error observed are when these texts are mixed into the body text around page breaks, interrupting reading flow. These types of errors are also observed frequently during screen reader use when reading directly from an untagged PDF. For section headings, in particular, the majority of papers have errors; around 67\% of papers have between 1--5 errors (either missed headings or extraneous headings), and 9\% have more than 5 errors. Due to the large number of section headings in papers, parsing errors are more frequent, and unfortunately, these errors impact the ability to properly navigate the HTML parse. 
Errors in body text extraction also negatively impact readability, in this case, select text in the document is being missed completely in the HTML render. We see that though the majority of parses have no body text errors, around 33\% of papers have between 1-5 missing paragraphs. 

Figure~\ref{fig:eval_results}(d) shows grading results for bibliography elements. Our pipeline is quite good at extracting bibliography entries, extracting all or most entries in the vast majority of cases, and successfully linking inline citations to these bibliography entries also in a large majority of cases. When bibliography extraction fails, it tends to fail catastrophically, resulting in no or few extractions.

The overall readability score is provided in Figure~\ref{fig:eval_results}(e). A majority of papers (54.5\%, 210 papers) have no major problems impacting readability. Another 31.7\% (122) of papers have some problems impacting readability, and 13.8\% (53) of papers have lots of readability problems. We are encouraged that a majority of HTML renders have no major problems, though our results necessitate further understanding of the papers with which our extraction pipeline has difficulty. If papers with lots of problems can be identified \textit{a priori}, we can prevent surfacing these low quality parses to the user. We perform some preliminary experiments to identify paper features that are more correlated with readability problems, though no features stood out as being predictive; we present those results in Appendix~\ref{app:eval_association}.

In Figure~\ref{fig:eval_fos}, we show the breakdown of overall readability by field of study,
plotting the proportion of papers per field that are classified as having no major problems, some problems, and lots of problems impacting readability. Many fields have similar distributions compared to the overall evaluation set. However, we note that some fields such as Art, Business, Economics, and Environmental science to some degree, have significantly lower quality extraction results. We posit that this may be due to biases in our PDF extraction pipeline. Some of the machine learning modules we use are primarily trained on paper data from the biomedical and Computer Science domains, where large scale labeled PDF extraction datasets can be found. Humanities-adjacent fields like Art and Business have very different publication norms, and the different layouts and content of papers and documents in these fields may provide additional challenges to our system, resulting in lower quality extraction and rendering.

\section{User study}
\label{sec:user_study}

We conduct an exploratory user study to better understand the needs of BLV scientists when reading papers, and to assess whether our prototype supports these needs. The study consists of a preliminary questionnaire and semi-structured video interview. Interviews are conducted remotely on Zoom.\footnote{\href{https://zoom.us/}{https://zoom.us/}} All recruitment materials, questionnaires, and the interview plan are reviewed and approved by the internal review board at \allenai. We recruit and interview \numusers users, with a pilot involving two users, and a main study involving four users. Modifications to the prototype between pilots and the main study can be found in Section~\ref{sec:pdf2html}. We report results from all \numusers participants in any analysis that does not involve the prototype, and for analysis that directly involves the prototype, we denote all cases where prototype modifications between the pilot and main study may impact our results.

The inclusion criteria for participants are:

\begin{itemize}
    \item The participant is over 18 years of age;
    \item The participant identifies as blind or low vision;
    \item The participant reads scientific papers regularly (more than 5 per year);
    \item The participant must have used a screen reader to read a paper in the last year; and
    \item The participant must complete the pre-interview questionnaire.
\end{itemize}

Participants were
recruited through mailing lists, word-of-mouth, and snowball sampling. Prior to each interview, the participant was asked to provide several keywords corresponding to their subject areas of interest, and between 3--5 papers where they experienced difficulty reading the PDF. Among the 3--5 papers, we selected one paper to use for the study, based on the availability of an HTML render, and maximizing the features that would be seen during the user study (e.g., given a choice between a paper with figures and a paper without figures but where both otherwise demonstrate the same paper components, we would select the paper with figures). Each study session was 75 minutes, consisting of three phases:

\begin{itemize}[itemsep=5pt]
    \item[] \textbf{Phase I: Capturing challenges with current work flow} \newline
    The primary research questions we investigate in this phase are: 
    \begin{itemize}[noitemsep, leftmargin=0.4in]
        \item[--] What methods and/or tools do BLV researchers use to assist in reading the literature?
        \item[--] What main accessibility challenges do BLV researchers face?
        \item[--] How do BLV researchers cope with these challenges?
    \end{itemize}
    We first asked the participant to describe their current workflow and the challenges they face when reading papers, clarifying how the user copes with challenges when their workflow does not adequately address the problem. We then asked the participant to demonstrate how they currently read a paper, by opening a paper PDF and walking us through the usage of their tools (PDF viewer, screen reader, magnifier, speech-to-text, etc). Participants kept their computer audio on so we could hear the output of their reader tools. The participant was asked to think aloud and describe their actions when reading the paper. We asked the participant to demonstrate any reading challenges they described in their pre-interview questionnaire. At the end of this phase, we asked the participant to assess how easy or difficult it was to read the paper with their current reading pipeline.
    \item[] \textbf{Phase II: Interaction with prototype} \newline 
    The primary research questions we investigate in this phase are: 
    \begin{itemize}[noitemsep, leftmargin=0.4in]
        \item[--] What features of the HTML render resonated positively with the participant?
        \item[--] What problems can be identified in the HTML render?
    \end{itemize}
    The goal of this phase was to understand how helpful or not helpful the HTML render is to the participant. The participant was asked to interact with an HTML render of the same paper they read in Phase I in the \scially prototype. We first provided an introduction to the prototype, then allowed the participant to proceed uninterrupted for several minutes interacting with the render. The participant was asked to think aloud during their interactions. Towards the end of this phase, we prompted the participant to interact with any features in the HTML render they may have skipped over. At the end of this phase, we asked the participant to assess how easy or difficult it was to read the paper with the HTML render.
    \item[] \textbf{Phase III: Q\&A and discussion} \newline 
    The primary objectives of this phase are to answer the questions:
    \begin{itemize}[noitemsep, leftmargin=0.4in]
        \item[--] How likely is the participant to use the HTML render in the future?
        \item[--] How can the HTML render be improved to best meet the participant's needs moving forward?
    \end{itemize}
    The participant was given further opportunities to ask questions or discuss the prototype. The participant was asked to describe their perceived pros and cons of the prototype, and to provide suggestions of missing features, ordered by priority. We asked the participant whether they would use this prototype if it were available, and if not, what features would need to be implemented to change that decision. 
\end{itemize}

\noindent The interviews were conducted by one author, with two other authors observing and participating during Phase III. All interviews were recorded for followup analysis, and participants were compensated with a \$150 USD gift card for their time. The questions used to guide the semi-structured interview are provided in Appendix~\ref{app:interview_questions}. 

We follow a grounded theory approach to identify themes and concepts from the participant interviews. We first perform open coding to identify relevant concepts, then axial coding to group these concepts under broad themes. These themes are 1) the technologies employed by users, 2) challenges in their current reading pipeline, and 3) mitigation or coping strategies, and in relation to the \scially prototype: 4) positive features, 5) negative features or issues with the prototype, and 6) suggestions for improvement. 
Interviews are selectively coded a second time to identify all concepts falling under each theme. We also employ the same method to code issues raised by participants in the pre-interview questionnaire. 

Themes and concepts are arrived upon by two authors following detailed reading of the interviews. In several cases, we further define attributes associated with some concepts, such as defining whether the technologies used were in relation to opening PDFs, screen reading, or other tasks; or whether the challenges identified affect the whole document, navigation, text, or a particular in-paper element. These delineations are described further in their respective results sections.

\subsection{Study participants}

Participants are graduate students, PhD students, and faculty members from predominantly English-speaking countries, whose primary research areas are in computer science, though also spanning neuroscience and mathematics. We interviewed two participants during the pilot phase and four participants during the main phase of our study. We report findings from all six participants for all themes captured in Phase I of the study. Since only minor changes were made to the prototype between the pilot and main study, we report findings from all participants for Phase II and III as well, making note of features that changed following the pilots. Three of six participants
study human-computer interaction and accessibility, which may be due in part to our sampling methodology, but may also reflect the relevance of accessibility research to BLV researchers. Other study participants conduct research in the areas of machine learning, neuroscience, software engineering, and blockchain. All but one participant reported having more than one year of experience using screen readers. The tools employed by participants are summarized in Table~\ref{tab:user_summary} along with the version of the \scially prototype with which they interacted.

\begin{table}[t!]
    \small
    \centering
    \begin{tabularx}{0.8\linewidth}{lllL{70mm}}
        \toprule
        \textbf{ID} & \textbf{Study} & \textbf{Prototype Version} & \textbf{Current Tools} \\
        \midrule
        P1 & Pilot & v0.1 & NVDA Screen Reader, Adobe Acrobat Reader \\
        \midrule
        P2* & Pilot & v0.2 & Mac Text-to-speech, Mac Magnifying Glass (sighted navigation), Mac Preview \\
        \midrule
        P3 & Main & v0.3 & Braille display, Mac VoiceOver, JAWS/NVDA on Windows, Mac Preview, Adobe Acrobat Reader \\
        \midrule
        P4 & Main & v0.3 & Mac VoiceOver, Mac Preview or Adobe Acrobat Reader \\
        \midrule
        P5 & Main & v0.3 & Microsoft Narrator, Adobe Acrobat Reader \\
        \midrule
        P6 & Main & v0.3 & Braille display, InftyReader, Mac VoiceOver, Mac Preview  \\
        \bottomrule
    \end{tabularx}
    \caption{User study participants, the prototype versions they interacted with, and the tools they currently use for reading papers. *P2 is low vision and uses sighted navigation tools in conjunction with a screen reader.
    }
    \label{tab:user_summary}
\end{table}

\subsection{Study findings}

\subsubsection*{Summary of current experience} 

Of the \numusers participants, three users have experience with screen readers on the Windows OS, such as NVDA, JAWS, and Microsoft Narrator, and three users use VoiceOver on MacOS. Two users use braille display in conjunction with their screen reader. One participant (P2) is low vision and uses a combination of text-to-speech and a magnifying glass to perform sighted navigation; P2's primary reading interaction involves selecting blocks of text in the PDF and using text-to-speech. Adobe Acrobat Reader is the most common software for opening PDFs; though several participants use Preview in MacOS, with one participant (P4) explicitly stating a preference for Preview over Acrobat. One participant uses a proprietary tool called InftyReader, which converts PDFs into ASCII text and math formulas into MathML, which is accessible.

\subsubsection*{Challenges of current PDF reading pipeline}

Table~\ref{tab:current_challenges} lists the challenges recognized by all participants in their current PDF reading pipeline. Some of these challenges affect the entire document, e.g., when a document lacks heading markup, it affects the ability to navigate the whole document. Others pertain to specific elements in PDFs, like inaccessible math formulas or lack of figure alt-text. All six users discussed the inaccessibility of math formulas. Unfortunately, document elements like math, figures, tables, and algorithm blocks are used to convey a significant amount of the information content of a paper, and the inability to access their content can produce negative impacts on the reader's ability to understand the paper.

\begin{table}[t!]
    \centering
    \begin{tabular}{lll}
        \toprule
        \textbf{Issue description} & \textbf{Affects} & \textbf{Raised by user} \\
        \midrule
        Scanned PDFs cannot be read without remediation & Document & P3, P4, P5* \\
        \midrule
        No headings/sub-headings for navigation & Navigation & P1, P3, P5 \\
        Figures are not annotated as figures & Navigation & P1, P5 \\
        Losing cursor focus when switching away from the PDF & Navigation & P1 \\
        Headings are not hierarchical (no sub-headings) & Navigation & P5 \\
        \midrule
        Text is read as single string (no spaces or punctuation) & Text & P1, P4, P5 \\
        Headers/footers/footnotes mixed into text & Text & P1, P4, P5 \\
        Words with ligatures are mispronounced & Text & P1, P3 \\
        Words split at line breaks are mispronounced & Text & P2, P3 \\
        Reading order is incorrect & Text & P3, P5 \\
        Text before and after figures sometimes skipped & Text & P4 \\
        Text on some pages not recognized at all & Text & P4 \\
        \midrule
        Math content is inaccessible & Element & P1, P2, P3, P4, P5, P6 \\
        Tables are inaccessible & Element & P1, P2*, P3, P5, P6 \\
        Figures lack alt-text & Element & P1, P3, P5, P6 \\
        Figure captions are not associated with figures & Element & P1, P5 \\
        Characters or words in figures are read and do not make sense & Element & P4, P5 \\
        Figure alt-text (when provided) is not descriptive & Element & P5 \\
        Code blocks are inaccessible & Element & P2, P4 \\
        \bottomrule
    \end{tabular}
    \caption{Challenges to PDF reading identified by participants during interviews. *Only identified as an issue during pre-interview questionnaire.}
    \label{tab:current_challenges}
\end{table}

\subsubsection*{Coping mechanisms}

The coping mechanisms employed by BLV researchers to read inaccessible PDFs are wide-ranging, often involving trying tools outside of their primary workflow, soliciting help from others, or in the worst case, giving up and moving on. We describe these in Table~\ref{tab:coping_mechanisms}. Several users reported trying certain tools like alternate PDF readers, browsers, or optical character recognition (OCR), even though the tools usually do not result in a significant improvement over their standard pipeline; when asked why, several participants reported feeling ``hopeful'' that a tool might work (P1) or hoping to get lucky (P3).

Several of these coping mechanisms involved other people. For example, three participants reported needing to ask sighted colleagues or family members to copy text, or to explain select paper content, especially figures and equations. Asking for PDF remediation was also a possibility for several participants; in this process, workers at the researcher's host institution convert a PDF into an accessible format, manually correcting equation representation and writing descriptions for figures. The output of the remediation process is seen as ``ideal'' (P4), but the process takes significant time (several weeks for any PDF) and may not fit into a researcher's schedule and timeline. Additionally, this process may only be available to researchers affiliated with a significantly large and resourced institution, and as P6 discusses, may no longer be a viable option for those who work outside of academia. In some cases, BLV researchers may also message authors directly to gain access to the source documents (P3 and P4). Both LaTeX source and Word documents are more accessible than PDFs, and access to these source documents can greatly improve the ability to read these papers.

Perhaps most disheartening is how often BLV researchers may simply give up in the face of an inaccessible paper. P1 says that by the time he has spent several hours making a paper readable, he may have already lost interest and motivation to read it. When asked how often papers are abandoned, P3 responds 60--70\% of the time. Though P4 does not discuss abandonment directly, P4 shares the following relevant sentiment: ``reading papers is the hardest part of research'' for a BLV researcher, and if papers were more accessible, there would be more blind researchers.

\begin{table}[t!]
    \small
    \centering
    \begin{tabular}{llp{60mm}}
        \toprule
        \textbf{Coping mechanism} & \textbf{Raised by user} & \textbf{What users said} \\
        \midrule
        Give up, abandon the paper & P1, P3, P5 & P3: when asked how often they abandon papers, answers ``60--70\% of the time'' \newline P5: sometimes the only option is to ``sit down and start crying'' (jokingly, though the sentiment is true) \\
        \midrule
        Try other conversion tools & P1, P3, P6 & \\
        \midrule
        Download LaTeX source or Word document if available & P3, P4, P6 & \\
        \midrule
        Ask sighted colleagues or family members to read & P3, P5, P6 & \\
        \midrule
        Ask for remediation / convert to braille & P4, P5, P6 & P4: 10 day turnaround is on the quick side, which is not good enough for research \newline P5: process takes a long time, around 1-2 weeks \\
        \midrule
        Try other PDF readers or browsers & P1, P6 & P1: may try Microsoft Edge browser even though it usually does not help, but he feels ``hopeful'' \\
        \midrule
        Message authors to get source document & P3, P4 & P4: sometimes the author manuscript is accessible but the camera-ready version is not; fault of the conferences and publishers, not the authors \\
        \bottomrule
    \end{tabular}
    \caption{Coping mechanisms discussed by users for dealing with challenging papers.}
    \label{tab:coping_mechanisms}
    \Description{
    }
\end{table}

\subsubsection*{Response to HTML render}

All user interviews were analyzed to extract positive and negative responses to various features or flaws of the prototype. We summarize these features and flaws in Table~\ref{tab:prototype_features}. Among the participants' favorite features are links between inline citations and references (all 6 participants), section headings for navigation (5 participants), the table of contents (4 participants), and figures tagged as figures with associated figure captions (3 participants). Regarding links between inline citations and references, several participants were especially supportive of the return links that allow the reader to return back to their reading context after following a citation link. P3 said that the links acted as external memory, allowing BLV users to essentially ``glance'' at the bibliography and back, like a sighted user might. Similar sentiments were shared by P5 and P6, although P5 also proposed the possibility of preserving the context even further by providing bibliography information inline rather than navigating back and forth between the main text and references section.

Among the negative features observed by participants, most have to do with imperfect extraction, for example, incorrectly extracted headings (3 participants), missed headings (2 participants), and various extraction issues with code blocks, tables, equations, and more. Many of these issues are known and quantified in Section~\ref{sec:evaluation}. Of these issues, problems with heading extraction were most notable, likely because the heading structure is the first element of the document with which the participants interact, and it provides a mental model of the overall document structure. Mistakes in heading extraction are obvious and erode trust in our overall system. As P5 says, ``it's really important that I trust it,'' and errors of this nature, both false positive and false negative extractions, can reduce trust. Similarly, though we describe in our introductory material that our system currently does not extract equations, P6 points out that it is unclear whether the system extracts equations because occasionally math can be found in the body text. This type of conflict between what is described and what is seen can also reduce trust. However, one may be able to build trust even in the face of extraction errors by indicating to the user when content is not extracted; as P4 says regarding the placeholders for not extracted items, ``at least I know there was an equation here.''

\begin{table}[t!]
    \footnotesize
    \centering
    \begin{tabularx}{\linewidth}{llp{60mm}}
        \toprule
        \textbf{Feature} & \textbf{Raised by user} & \textbf{What users said} \\
        \midrule
        \textsc{Positive} & & \\
        \midrule
        Bidirectional links between inline citations and references & P1, P2, P3, P4, P5, P6 & P3: ``very few research teams actually get this and get this right, so well done''; ``crucial piece of the puzzle'' \\
        Headings for easy navigation & P1, P2, P3, P4, P6 & P4: ``Headings are the best thing ever''; makes it very clear what section you are in \\
        Table of contents* & P2, P3, P5, P6 & \\
        Figures are tagged as figures, and captions are associated & P4, P5, P6 & \\
        Can use browser and OS features like find/copy/paste  & P1, P4 & \\
        Simple typography for reading & P2 & \\
        Can interact with headings word-by-word or letter-by-letter & P4 & \\
        Not extracted items are noted as missing & P4 & P4: ``at least I know there was an equation here'' \\
        \midrule
        \textsc{Negative} & & \\
        \midrule
        Some headings extracted incorrectly & P1, P3, P5 & \\
        Some headings missed in extraction & P3, P5 & P5: ``it's really important that i trust it''; ``there [should be] *no* false negatives'' \\
        Code block not extracted & P2, P4 & \\
        Tables are extracted as figures & P2, P6 & \\
        Equations not extracted & P4, P6 & P6: Not sure if this system extracts equations because sometimes there is some math in the body text \\
        Figures placed away from text* & P1 & \\
        No alt-text extracted & P1 & \\
        URLs missing from bibliography entries** & P2 & \\
        Some information not surfaced (keywords, footnotes) & P3 & \\
        Some headers/footers/footnotes mixed in text & P4 & \\
        Headings are not hierarchical & P5 & \\
        \bottomrule
    \end{tabularx}
    \caption{Positive and negative features identified in the prototype. *The feature was implemented or the issue addressed in v0.2 following P1 pilot. **The issue was addressed in v0.3 following P2 pilot.}
    \label{tab:prototype_features}
\end{table}

\subsubsection*{Difficulty scale} 

The responses of the users to the difficulty of their current pipeline versus the HTML render are shown in Table~\ref{tab:taskload}. We ask the following question: \textit{On a scale of 1 to 5, how easy or difficult was it to read this paper with the HTML render, and why? (Answers: 1 = Very easy; 2 = Easy; 3 = Neutral; 4 = Difficult; 5 = Very difficult)}

All participants in the main study reported that the HTML render is easier for reading than their current pipeline. Reductions in difficulty rating ranged from 0.5 to 3.0. Most of our participants rated their current pipeline as difficult (4 participants) or neutral (1 participant), with one participant who is low vision (P2) reporting that their current pipeline is easy. During our pilot sessions, users reported that the HTML render was difficult to use. For the main study, users reported the HTML render as neutral or easy to use.

P2 is the only participant to report the HTML render as being more difficult to use than their current pipeline; we note that P2 is sighted and did not engage with most of the navigation features we designed and implemented for screen reader-based navigation. Because P2 primarily interacted with papers through sighted navigation, text highlighting, and text-to-speech, they were able to interact with section headers, figures, tables, and equations in the original PDF using the magnifier tool, and found any missing content in the HTML render to be significantly detrimental to their reading experience.

The overall median difference in difficulty scores between the PDF and HTML render is modest, at 0.75. This modest change may be due to the conflation of interface design and system errors when asking participants to rate the difficulty of use. In general, all users responded very positively to the interface design, especially around the navigational features we introduce. Issues were raised around extraction accuracy and the propagation of these errors to the interface. We may be able to offset some of the latter issues by detecting and removing papers that suffer from more extraction errors, though we leave this to future work.

\begin{table}[tb!]
    \centering
    \begin{tabular}{llp{12mm}p{12mm}lp{15mm}}
    \toprule
        \textbf{ID} & \textbf{Study} & \textbf{Current pipeline} & \textbf{HTML render} & \textbf{Difference} & \textbf{Would use in future} \\
    \midrule
        P1 & Pilot & 4.0 & 4.0 & 0.0 & Yes \\
        P2 & Pilot & 2.0 & 4.0 & \color{red}{-2.0} & Yes \\
    \midrule
        P3 & Main & 3.0 & 2.0 & \color{darkgreen}{1.0} & Yes \\
        P4 & Main & 4.0 & 1.0 & \color{darkgreen}{3.0} & Yes \\
        P5 & Main & 4.0 & 3.0 & \color{darkgreen}{1.0} & Yes* \\
        P6 & Main &4.0 & 3.5 & \color{darkgreen}{0.5} & Yes \\
    \bottomrule
    \end{tabular}
    \caption{Participant ratings on the difficulty scale (1 = very easy, 2 = easy, 3 = neutral, 4 = difficult, 5 = very difficult) and whether they would use the tool in the future. All participants reported a change from more difficult to more easy when moving from their current pipeline to the HTML render except P2, who uses sighted navigation. The median reduction in difficulty score for all participants is 0.75. All participants reported that they would be very likely to use the system in the future were it to be available; P5's response is contingent on improvements in section heading extraction.
    }
    \label{tab:taskload}
\end{table}

\subsubsection*{Future usage}

At the end of each session, we ask users whether they would be likely to use the prototype in the future if it were made publicly available on a range of papers. We ask specifically: \textit{On a scale of 1 to 5, how likely are you to use the HTML render, if it is available to you in the future? (Answers: 1 = Very unlikely, 2 = Unlikely, 3 = Neutral, 4 = Likely, 5 = Very likely)} If the answer is unlikely or neutral, we ask what changes would need to be made to the tool such that they would use it.

All users reported that they would use the prototype in the future.
Five users responded 5, that they would be very likely to use it; one user (P5) responded 3 to the prototype as it currently is, and 5 if some of the issues for heading extraction were addressed. P1, who participated in an early pilot with fewer implemented features, said that this would become a tool in the toolbox, but he would not be able to rely solely on it due to incomplete extractions. P5 expressed a similar sentiment, that in its current state, he may try the prototype system when his current workflow fails, but if issues around heading extraction were addressed, he would be very likely to use it. P3 replies when asked how the system might be integrated into their workflow, ``I think it would become the workflow.'' P4 says ``for unaccessible PDFs, this is life-changing.''

\subsection{Design recommendations}
\label{sec:designrecs}

\begin{figure}[t!]
    \centering
    \includegraphics[width=0.8\linewidth]{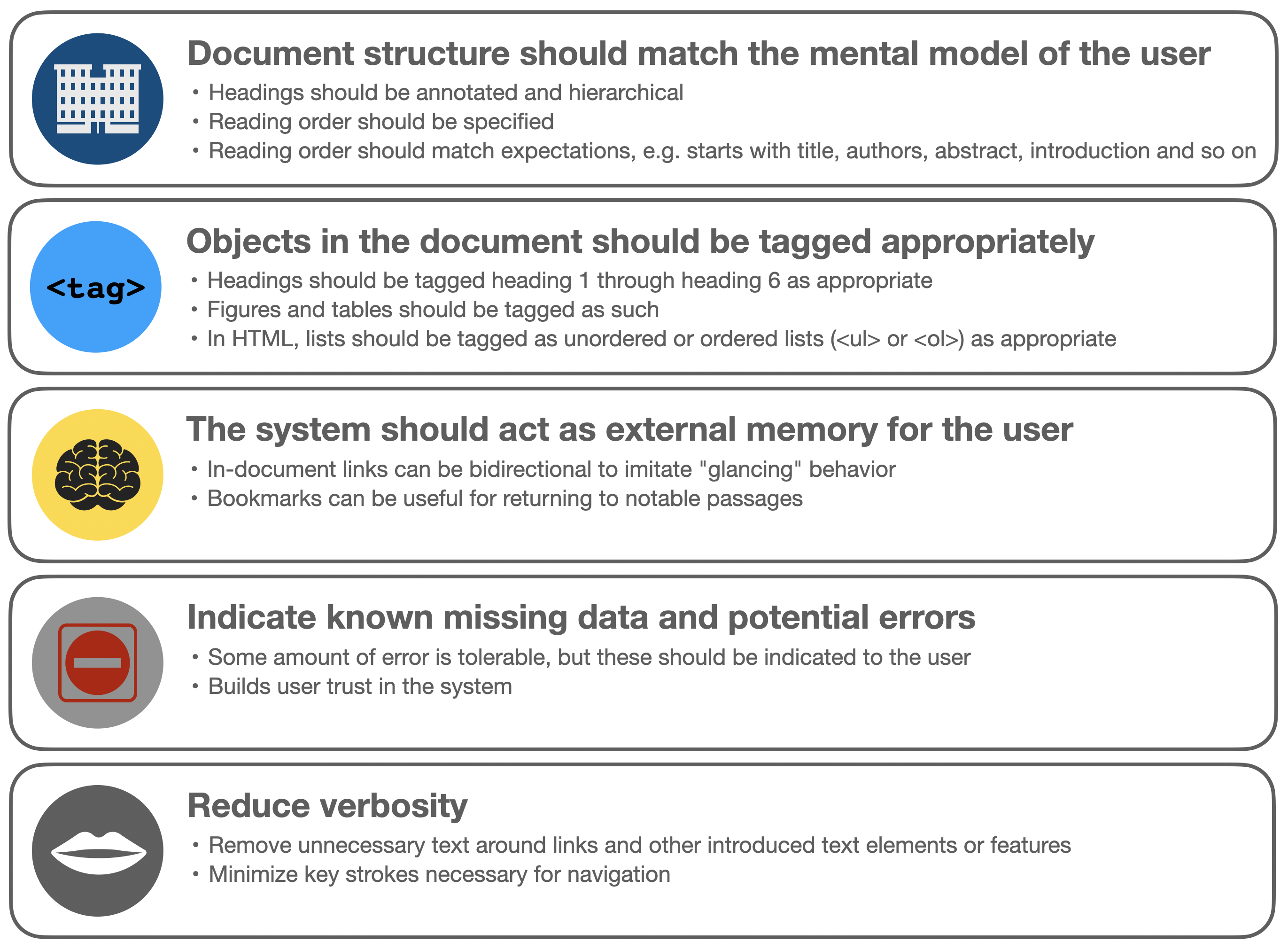}
    \caption{Design recommendations for screen reader friendly paper reading systems. A system should aim to provide the document structure in a way that matches the mental model of the user, and to tag all elements appropriately. These aspects are achievable through proper tagging of a paper, including in PDF format. Additionally, a system should aim to act as external memory for the user, minimizing the amount of cognitive load needed to return to their reading context. To improve trust, a system should indicate when there is known missing data in the extraction or a possibility of missing or incorrect data. Finally, a system should reduce verbosity, ensuring that as few keystrokes as possible are necessary for the user to perform their desired task. 
    }
    \label{fig:design_recs}
    \Description{Five design recommendations. 1. A picture of a building on a circular dark blue background. Text that reads "Document structure should match the mental model of the user; headings should be annotated and hierarchical; reading order should be specified; reading order should match expectations, e.g. starts with title, authors, abstract, introduction and so on." 2. A picture of the word <tag/> on a circular blue background. Text that reads "Objects in the document should be tagged appropriately; headings should be tagged heading 1 through heading 6 as appropriate; figures and tables should be tagged as such; in HTML, lists should be tagged as unordered or ordered lists (<ul> or <ol>) as appropriate." 3. A picture of a brain on a circular yellow background. Text that reads "The system should act as external memory for the user; in-document links can be bidirectional to imitate 'glancing' behavior; bookmarks can be useful for returning to notable passages." 4. A picture of a no entry sign on a circular light gray background. Text that reads "Indicate known missing data and potential errors; some amount of error is tolerable, but these should be indicated to the user; builds user trust in the system." 5. A picture of a pair of lips on a circular dark gray background. Text that reads "Reduce verbosity; remove unnecessary text around links and other introduced text elements or features; minimize key strokes necessary for navigation."}
\end{figure}

We distill our learnings into a set of five design recommendations for BLV user-friendly paper reading systems. Figure~\ref{fig:design_recs} summarizes the following recommendations:

\begin{enumerate}[itemsep=4pt]
    \item[1.] \textbf{Document structure should match the mental model of the user.} Structure is necessary for providing an overview of a document and is essential to navigation. Headings in a paper should be tagged as such and the hierarchy of the headings should match the mental model of the user, i.e., top level headings should be tagged \texttt{<h1>} or \texttt{<h2>}, and lower level headings \texttt{<h3>} through \texttt{<h6>} accordingly. Reading order should be specified, as to not interject non-body text objects into the body text, e.g., headers, footers, and footnotes often disrupt the main flow of text because they visually break paragraphs. Similarly, a user expects a natural flow to a paper, beginning with the title, authors, abstract, introduction etc, and ending with conclusions and references. Papers with various elements interspersed are disruptive of this mental model and can interfere with the reader's understanding of the document.

    \item[2.] \textbf{Objects in the paper should be tagged appropriately.} Self-explanatory. Headings should be tagged as headings, figures as figures, tables as tables, lists as lists and so on. Appropriate tagging allows a user to take advantage of the screen reader's capabilities for navigating to specific types of objects, e.g., most screen readers have shortcuts for navigating headings, and to figures or lists. Proper tagging emulates a sighted user's ability to detect visually distinct objects such as headings, figures, and tables. When objects are not appropriately tagged, a screen reader user must scroll through the whole document each time to identify the desired sections.

    \item[3.] \textbf{The system should act as external memory for the user.} 
    Visual layout can act as a source of external memory for sighted users, who can quickly derive reading context and object types from visual cues. For BLV users, strategies for emulating such external memory can be beneficial. For example, bi-directional navigation for all in-document links are a type of ``glancing'' feature. With this feature, a user no longer needs to commit text to memory in order to rediscover their previous reading context after navigating away. P3, in particular, emphasizes that these features are a ``crucial piece of the puzzle.'' Other memory features like bookmarking or note-taking may also be helpful for returning the user to their reading context.

    \item[4.] \textbf{Indicate known missing data and potential errors.} To facilitate trust in the system, the system should indicate the presence of missing and erroneous data to users. Some degree of fault tolerance is permitted, as long as the overall benefit to the user is greater. However, as these systems rely on statistical methods, extraction quality is rarely perfect. Most users indicate a preference for knowing when the system fails, rather than dealing with the uncertainty of figuring out whether the issue is with the underlying paper, or with the extraction and reading interface.

    \item[5.] \textbf{Reduce verbosity.} Any minimization of unnecessary text and spaces between links can simplify navigation for BLV users. Though these extra commas and spaces may seem innocuous for sighted users, they require extra keystrokes for screen readers. Reduction of unnecessary verbosity around links and introduced features can save time for screen reader users.

\end{enumerate}

The overarching themes of these recommendations are to reduce user cognitive load and improve trust in the system. Regarding cognitive load, interruptions to reading flow for BLV users are especially disruptive, since there are no visual markers to help identify reading context. Paper reading systems for BLV users should therefore attempt to mitigate cognitive load caused by loss of context, by allowing users to quickly navigate back to their reading context when following any links, and by avoiding any disruption of reading flow. Regarding this latter point, properly labeled reading order, headings for navigation, and appropriately tagged objects all contribute to mitigating disruptions. Further, it is also important to remove interjections from headers, footers, footnotes, figure and table captions, and other text, all of which interrupt the natural flow of reading.

Regarding user trust in the system: this should a priority of any system builder. Because PDF extraction and document rendering are imperfect processes, some degree of error is expected. Though all participants in our user study expressed that some degree of error is tolerable, one can mitigate the conversion of errors to distrust by clearly indicating known errors and missing content in the system. For example, in some cases our system is unable to extract a figure caption; if the caption for Figure 3 is not extracted, rather than skipping from Figure 2 to Figure 4 and causing confusion for the reader, it is better to indicate that Figure 3 is missing in the extraction.

A system that responds quickly to user requests is obviously more desirable. However, several participants indicated that some wait time is acceptable, especially if a longer wait time corresponds to a higher quality reading experience. Though we report this finding, we ask readers to take it with a grain of salt. This point may not hold for all or even a majority of users, since several users also remark on the PDF remediation process (which usually takes 1--2 weeks) as being too long to adequately support their research workflow.

Though we derive these design recommendations in the scope of paper reading, they are generalizable to other classes of documents. In fact, several of these design principles echo available guidelines for human-AI interaction \citep{Amershi2019GuidelinesFH}, especially in indicating the capabilities and limitations of the system (recommendation 4). A number of our recommendations are simply good practice, such as exposing the structure of a document and tagging document objects appropriately, and are covered by current guidelines for creating accessible documents. Other recommendations focus on emulating the types of advantages that sighted users derive from layout and visual information, but to implement them in such a way that BLV users can benefit, e.g. using the system as a source of external memory.

\section{Discussion}
\label{sec:discussion}

In this work, we present the results of several studies that aim to characterize the current state of accessibility for academic paper PDFs, to learn the challenges faced by BLV researchers when reading papers, and to demonstrate how our \scially system that renders PDFs into accessible HTML can be used to mitigate many of these challenges. 

Based on our analysis, the current state of paper accessibility is grim, with an average of \percaccessible of papers across all fields of study satisfying our five assessed accessibility criteria. Though there is some improvement seen over time, we are not optimistic that these improvements are due to authors prioritizing accessibility when writing papers, since the presence of figure alt-text (the only of the five criteria that requires author intervention) remains low. Rather, the commitment to accessibility made by certain typesetting software providers such as Microsoft Word may be responsible for a portion of these improvements. Given the strong correlation between PDF creation software and accessibility compliance, we encourage conferences, publishers, and authors to consider the tools they are using to generate PDFs, and to integrate accessibility requirements during the publication process.

Given the scope and magnitude of the problem, and how PDF is still the dominant file type used for distributing scientific papers, there are clear needs for immediate technological solutions. We propose the \scially system, which integrates several text and vision machine learning models to extract the content from paper PDFs and render this content as HTML. The system adds tags and infers reading order, thereby improving the navigational capabilities of BLV users. Of course, no extractive pipeline is perfect, and we quantify and qualify extraction quality through an evaluation study and user study. Our intrinsic evaluation of extraction quality indicates that most extractions have no major problems affecting readability (86.2\% have no or only some problems). The most common extraction problems are incorrectly extracted or missed section headings, as well as headers, footers, and footnotes being improperly mixed into the body text, which can interrupt reading flow. Participants in our user study responded positively to \scially, preferring its navigational features and tagging to working with PDFs. Though the various types of extraction mistakes made by our system are noted by participants, most participants reported an improvement from their current reading pipeline, and all participants expressed an interest in using the system in the future. 

We present the challenges, coping mechanisms, and positive and negative features identified by participants. We also summarize the collective themes into a set of five design recommendations for other researchers and practitioners looking to design and build systems for accessible reading. The recommendations include (1) matching the document structure to the mental model of the user, (2) tagging all objects within the document appropriately, (3) acting as external memory for the user, (4) indicating known missing data or extraction errors, and (5) reducing verbosity. The first two of these recommendations are related to proper and correct representation of the document structure and in-paper objects. Both are necessary components of an accessible document. The third recommendation is to provide additional navigation features that are otherwise encoded in the visual layout of the document and inaccessible to BLV users. The fourth recommendation is related to error tolerance and user trust. For any machine learning-based document parsing system, errors are inevitable; managing user expectations for these systems is crucial. This recommendation echoes previously published guidelines for human-AI interaction, which suggest communicating to the user the capabilities and limitations of the AI system \citep{Amershi2019GuidelinesFH}. Setting expectations correctly and referring the user back to the original source document when the extractive procedure fails can help mitigate inappropriate reliance on the system. The final recommendation aims to reduce verbosity and the number of keystrokes needed for performing any task, which can speed up the use of such a system.

We hope these design recommendations will facilitate further conversations around the needs of BLV users, and that they may result in systems that ease the reading burden for these users. As one participant puts it, ``reading papers is the hardest part of research'' for researchers who are blind or low vision, and if papers were more accessible, ``there would be more blind researchers.'' It is a duty of the entire community to facilitate this, and to design, prototype, and build systems to support the needs of the BLV research community.

\subsection{Limitations \& Future Work}
\label{sec:future_work}

This work focuses on rendering PDF papers in HTML to improve document navigation and provide a more intuitive reading order. There are many other aspects of accessibility with which we do not contend, such as providing figure alt-text, accessible math, or tagging tables. Future work involves investigating various ways to improve or provide these features automatically, or by harnessing the power of the community to provide some of these features for papers as they are requested. For example, we may integrate element-specific reading features for mathematical equations \citep{Flores2010MathMLTA, Bates2010SpokenMU, Sorge2014TowardsMM, Mackowski2017MultimediaPF} or graphs and charts \citep{Elzer2008AccessibleBC, Engel2017TowardsAC, Engel2019SVGPlottAA}, or create a crowd-sourcing pipeline to solicit alt-text annotations for figures that lack descriptions.

PDF parsing remains an open research problem with many challenges. Our reliance on these technologies necessarily introduce error into our pipeline and system. We attempt to describe and quantify these errors in Appendix~\ref{app:eval_association}, but found no strong correlation between any particular type of error and the overall quality assessment. Unfortunately, this means that there is no obvious mitigation strategy for identifying low-quality extractions before they are shown to users. Further work remains to automatically or semi-automatically identify low-quality parses prior to surfacing them. For example, we could investigate other paper features as predictors of parse quality. With more labeled data, we could also train a neural classifier to identify low-quality parses. 

In this work, we focus on processing PDFs and making them accessible. Some papers are available in XML, HTML, or other structured markup languages; and LaTeX or Word document source can be found for others. Our system could take advantage of these alternatives to PDFs when they are publicly available, for example, by rendering the semantic content of the paper as extracted from these other document representations, as in arXiv Vanity\footnote{\href{http://www.arxiv-vanity.com/}{http://www.arxiv-vanity.com/}} for arXiv LaTeX source or Pubmed Central's PubReader,\footnote{\href{https://www.ncbi.nlm.nih.gov/pmc/about/pubreader/}{https://www.ncbi.nlm.nih.gov/pmc/about/pubreader/}} which renders JATS XML. Though S2ORC \citep{lo-wang-2020-s2orc} contains LaTeX parses derived from arXiv for over 1 million papers, further study is necessary to determine whether these parses are suitable for HTML rendering in our system.

Though we conduct a user study to better understand the challenges of BLV users and their responses to our prototype, the number of participants involved is small. Consequently, we focus on identifying qualitative learnings from these user studies. These learning, when combined with our evaluation and analysis of the current state of scholarly PDF accessibility, provide a more complete portrait of the challenges and issues BLV scholars face when reading papers. To more fully assess the benefits and flaws of our system, a broader user study and testing period is needed. We hope to achieve this in future work.

Lastly, PDFs have been repeatedly called out as being inaccessible, not only for screen readers, but broadly for reading, especially on mobile and other devices with small screen sizes \citep{NielsenPDFStillUnfit}. Dissociating publishing from PDFs continues to be a good goal for the future. In recent years, alternative publication formats have risen in popularity, such as eLife's dual publication in PDF and HTML,\footnote{\href{https://reviewer.elifesciences.org/author-guide/post}{https://reviewer.elifesciences.org/author-guide/post}} the interactive HTML papers at distill.pub,\footnote{\href{https://distill.pub/}{https://distill.pub/}} or the ACM Digital Library's very own dual publication (PDF and HTML) process,\footnote{\href{https://www.acm.org/publications/authors/submissions}{https://www.acm.org/publications/authors/submissions}} which is now available for many of the ACM's computing conferences and journals. We have no doubt that viable alternatives to PDF have and will arise, and encourage the community to explore these options when making publication decisions.
\section{Conclusion}

Based on our findings, most academic papers are inaccessible and significant challenges remain for BLV researchers when interacting with and reading these papers. Though some improvements in accessibility have been seen over time, these changes may not be reflective of author actions directly. In the meantime, we offer a potential solution for the millions of PDFs that have already been published and which still remain the dominant form of distribution for academic papers. We introduce the \scially system for rendering PDFs as accessible HTML documents. The system extracts the content of PDFs, tagging headings and objects and inferring reading order, which results in a more navigable and accessible document. Though the extraction pipeline is imperfect and can result in errors, our evaluation suggests that for the majority of papers, the resulting HTML render has no major problems that impact readability. We confirm these findings in our user study, where all users responded positively to the prototype system, claiming that they would be likely or very likely to use the system were it to be available in the future. Participants described the system as likely to ``become the workflow'' or ``life-changing,'' indicating both a strong favorable response and particular need for these types of solutions.

We do not claim that \scially solves all (or even close to all) accessibility problems for BLV researchers, but it is a step in the right direction.  \scially is a technological solution that can mitigate many of the challenges experienced by BLV researchers at this moment. Though a longer term solution would surely require more dialogue between all stakeholders and a potential revolution in the way in which scholars publish and distribute their research findings, we encourage researchers to prioritize and address these challenges with whatever tools they have in their toolbox right now. We especially encourage others to take into account our findings on the needs and challenges of BLV researchers when designing and engineering new systems and tools for reading the scholarly literature.

\begin{acks}
This work was supported in part by ONR grant N00014-18-1-2193, NSF RAPID grant 2040196, and the University of Washington WRF/Cable Professorship. We thank Jeff Bigham, Leah Findlater, Jon Froehlich, and Venkatesh Potluri for their valuable feedback on study design and recruitment. We thank Oren Etzioni and Doug Raymond for valuable feedback on the project. We thank Bryan Newbold for providing feedback on earlier drafts of the manuscript. We thank Sam Skjonsberg for help with the demo, and Michal Guerquin and Michael Schmitz for feedback on demo deployment. We thank the Semantic Scholar team for assisting with data access and system infrastructure. Finally, we thank the users who participated in our study, who offered invaluable feedback and suggestions.
\end{acks}

\bibliographystyle{ACM-Reference-Format}
\bibliography{a11y}

\newpage
\appendix

\section{Evaluation forms}

This section contains forms and documents used to evaluate the quality of HTML renders produced by our system.

\subsection{Evaluation instructions}
\label{app:eval_instructions}

Instructions for annotators are reproduced verbatim below.

\begin{center}
\noindent \rule{0.9\linewidth}{0.4pt}
\end{center}
\begin{quote}
\textbf{Goal:} Identify and quantify the prevalence of different parse issues in S2ORC parses to assess their suitability for accessibility purposes. This will help us decide whether S2ORC parses can help meet screen reacher accessibility needs.

\vspace{4pt}\noindent\textbf{Number of papers:} ~500 papers sampled across different domains of science

\vspace{4pt}\noindent You will be presented with a spreadsheet of scientific papers, each with a pair of links. One link goes to a PDF of the paper. One link goes to an HTML representation of the same article. For each pair of links, we would like to know how faithfully the HTML representation captures the information on the PDF.

\vspace{4pt}\noindent\textbf{INSTRUCTIONS:}

\begin{enumerate}
    \item[1.] Open the two links side by side.
    \item[2.] If the two links do not seem to correspond to the same paper, STOP. Make a note in the spreadsheet and SKIP.
    \item[3.] If the PDF shows a paper that is not suitable, STOP. Make a note in the spreadsheet and SKIP. Non-suitable may include:
    \begin{itemize}
        \item[--] It is not a scientific paper.
        \item[--] It is spam or a fake paper.
        \item[--] It is slides, a poster, or other such non-paper document.
        \item[--] It is just an abstract.
        \item[--] It is a series of articles (e.g. conference proceedings, journal issue etc).
        \item[--] It is a book.
        \item[--] It is supplementary material; \\
        note: some supplementary material is solely made up of figure or images.
        \item[--] Something else that makes you pause. If you're not sure, SKIP it. 
    \end{itemize}
    \item[4.] Copy the paper identifier corresponding to this paper into the first question on this form. Please make sure the identifier matches the paper you are evaluating.
    \item[5.] Answer each of the questions in this form as best as you can, treating the PDF as gold. There is no need to review every word or line of text. We are just trying to get an overall assessment of parse quality. For any question that asks for a number, enter `0` if there are no obvious problems with those extractions.
    \item[6.] Submit the form and mark the row in the spreadsheet as complete.
\end{enumerate}

\vspace{4pt}\noindent\textbf{Note:} Display equations (those that are in their own paragraph) are currently not preserved in S2ORC, so we ask the annotator to ignore issues around missing display equations. Inline equations (those that are inline within a paragraph) are converted to token streams, which may not be faithful to the original PDF (e.g. fractions may not be preserved). The annotator can provide a description of issues around equation parsing when there is a notable issue.
\end{quote}
\begin{center}
\noindent \rule{0.9\linewidth}{0.4pt}
\end{center}

\subsection{Evaluation questions}

Questions asked in the evaluation form are reproduced in Table~\ref{tab:eval_questions}.

\begin{table}[h!]
\small
\begin{tabular}{lp{130mm}}
    \toprule
    \textbf{Answer} & \textbf{Questions} \\
    \midrule
    y/p/n & Is the TITLE correctly extracted? \\
    text & \textit{Comment (clarify if answer is ``partially'' or ``no'')} \\
    \midrule
    y/p/n & Are the AUTHOR(S) correctly extracted? \\
    text & \textit{Comment (clarify if answer is ``partially'' or ``no'')} \\
    \midrule
    y/p/n & Is the ABSTRACT correctly extracted? \\ 
    text & \textit{Comment (clarify if answer is ``partially'' or ``no'')} \\
    \midrule
    y/n & Does this paper contain a substantial number of math EQUATIONS (more than 5 display equations)? \\
    \midrule
    number & How many FIGURES are in the PDF? (Enter `0' if none) \\
    number & How many FIGURES are correctly extracted? (Enter `0' if no figures in paper) \\
    number & How many FIGURE CAPTIONS are correctly extracted? (Enter `0' if no figures in paper) \\
    number & Approximately how many FIGURE captions are **INCORRECTLY** parsed into the body text (should be a figure caption but is mixed in with the body text)? (Enter `0' if they are all correct or if no figures) \\
    text & \textit{Comment (Optional -- Note anything here about FIGURES or FIGURE CAPTIONS, e.g. which figures are not extracted, which figure captions are not extracted, which figure captions are incorrectly extracted into the body text etc.)} \\
    \midrule
    number & How many TABLES are in the PDF? (Enter `0' if none) \\
    number & How many TABLES are correctly extracted? (Enter `0' if no tables in paper) \\
    number & How many TABLE TITLES / CAPTIONS are correctly extracted? (Enter `0' if no tables in paper or if those tables do not have titles / captions) \\
    number & Approximately how many TABLE titles are **INCORRECTLY** parsed into the body text (should be a table title / caption but is mixed in with the body text)? (Enter `0' if they are all correct or if no tables or table titles / captions) \\
    number & Approximately how many TABLES have content that is **INCORRECTLY** parsed into the body text (content of table is mixed with the body text)? (Enter `0' if they are all correct or if no tables) \\
    text & \textit{Comment (Optional -- Note anything here about TABLES or TABLE TITLES / CAPTIONS, e.g. which tables are not extracted, which table captions are not extracted, which table title / captions / content are incorrectly extracted into the body text etc.)} \\
    \midrule
    number & Approximately how many times are page HEADERS or FOOTERS *INCORRECTLY* mixed into the body text?This also includes margin content such as arXiv watermarks. (Enter `0' if all okay or no headers or footers) \\
    text & \textit{Comment (Optional -- Note anything interesting here about incorrectly parsed headers or footers; no need to provide page numbers)} \\
    \midrule
    number & Approximately how many SECTION HEADINGS are *INCORRECTLY* extracted?  (Enter `0' if they are all correct or no section headings) \\
    text & \textit{Comment (Optional -- Note anything interesting about the section heading extractions, no need to list exhaustively)} \\
    \midrule
    number & Approximately how many BODY TEXT PARAGRAPHS are **MISSING** from the extraction?  (Enter `0' if they are all there or there is no body text) \\ 
    text & \textit{Comment (Optional -- Note anything interesting about the body text extractions)} \\
    \midrule 
    choice & Are BIBLIOGRAPHY entries extracted correctly? (options: all correct, mostly correct, half correct, mostly incorrect, incorrect, no bibliography)  \\
    text & \textit{Comment (Optional -- Note anything interesting about the bibliography extractions; no need to list exhaustively)} \\
    \midrule
    choice & Are INLINE CITATIONS linked to bibliography entries? (Please answer this questions considering only the bibliography entries that were extracted) (options: all linked, majority linked, half linked, most unlinked, none linked, no bibliography) \\
    text & \textit{Comment (Optional -- Note anything interesting about the inline citation linking; no need to list exhaustively)} \\
    \midrule 
    text & \textit{Are there any other problems with the HTML parse that are not covered by one of the above questions? Please describe. (Optional)} \\
    \midrule
    choice & Please rate the overall full text quality in the HTML render (options: no major problems, some problems, lots of problems -- see rubric in Section~\ref{app:quality_rubric}) \\
    \bottomrule
\end{tabular}
\caption{Evaluation questions. Optional questions are in \textit{italics}.}
\label{tab:eval_questions}
\end{table}

\subsection{Quality rubric}
\label{app:quality_rubric}

The quality rubric for the final question in the evaluation form is given in Table~\ref{tab:quality_rubric}. This rating attempts to capture the overall readability and usability of the HTML render. Three authors discussed and converged upon this rubric following initial pilot annotations.

\begin{table}[h!]
    \small
    \centering
    \begin{tabular}{M{30mm}m{80mm}}
        \toprule
        Rating & Criteria \\
        \midrule
        No major problems that impact readability &
        \begin{itemize}[leftmargin=*]
            \item No errors or relatively few errors
            \item No missing paragraphs, but a few insertions into paragraphs or incorrect headers okay
            \item Any errors impact only a couple of paragraphs
        \end{itemize} \\
        \midrule
        Some problems that impact readability & 
        \begin{itemize}[leftmargin=*]
            \item Few missing paragraphs (<1 per 5 pages) OR Several figure/table insertions into paragraphs or incorrect headers
            \item Errors can impact multiple paragraphs
        \end{itemize} \\
        \midrule
        Lots of problems that impact readability & 
        \begin{itemize}[leftmargin=*]
            \item Difficult to read
            \item Multiple missing paragraphs OR multiple figure/table insertions that make some paragraphs unreadable
            \item Errors impact majority of paragraphs
        \end{itemize} \\
    \bottomrule
    \end{tabular}
    \caption{Rubric for HTML parse quality assessment (final question in evaluation questionnaire).}
    \label{tab:quality_rubric}
\end{table}

\section{Evaluation results}
\label{app:eval_raw_results}

Raw counts for each type of error detected during the evaluation of HTML renders are provided in Table~\ref{tab:eval_raw_by_element}. The overall quality score split by field of study is shown in Table~\ref{tab:eval_raw_by_fos}.

\begin{table}[h!]
\small
    \centering
    \begin{tabularx}{\linewidth}{L{38mm}L{18mm}L{18mm}L{18mm}L{18mm}L{18mm}}
    \toprule
        \textbf{Metadata Element} & \textbf{Yes} & \textbf{Partially} & \textbf{No} & & \\
    \midrule
        Title & 337 & 16 & 32 & & \\
        Authors & 307 & 64 & 14 & & \\
        Abstract & 308 & 22 & 55 & & \\
    \midrule
        \textbf{Figure/Table Element} & \textbf{Skipped} & \textbf{No figures/tables} & \textbf{No errors} & \textbf{1 error} & \textbf{>1 error} \\
    \midrule
        Figure extraction errors & 6 & 94 & 201 & 45 & 39 \\
        Figure caption errors & 0 & 94 & 174 & 55 & 62 \\
        Table extraction errors & 2 & 166 & 165 & 32 & 20 \\
        Table caption errors & 2 & 166 & 190 & 23 & 4 \\
    \midrule
        \textbf{Text Element} & \textbf{Skipped} & \textbf{No errors} & \textbf{1-5 errors} & \textbf{>5 errors} & \\
    \midrule
        Header/Footer/Footnote errors & 3 & 170 & 172 & 40 & \\
        Section heading errors & 2 & 88 & 258 & 37 & \\
        Body paragarph errors & 1 & 226 & 128 & 30 & \\
    \midrule
        \textbf{Bibliography Element} & \textbf{Skipped/poor bib extraction} & \textbf{No bibliography} & \textbf{All or most correct} & \textbf{Half correct} & \textbf{Mostly incorrect} \\
    \midrule
        Bibliography extraction & 7 & 15 & 313 & 3 & 47 \\
        Inline citation linking & 39 & 10 & 290 & 20 & 26 \\
    \midrule
        \textbf{Overall Readability} & \textbf{Good} & \textbf{Okay} & \textbf{Bad} & & \\
    \midrule
        Overall score & 210 & 122 & 53 & & \\
    \bottomrule
    \end{tabularx}
    \caption{Assessment count for all evaluation paper elements. Corresponds to distributions shown in Figure~\ref{fig:eval_results}.}
    \label{tab:eval_raw_by_element}
\end{table}

\begin{table}[h!]
\small
    \centering
    \begin{tabularx}{0.73\linewidth}{L{32mm}L{16mm}L{16mm}L{16mm}X}
    \toprule
        \textbf{Overall Readability} & \textbf{Number of papers} & \textbf{Good} & \textbf{Okay} & \textbf{Bad} \\
    \midrule
        All papers & 385 & 210 & 122 & 53 \\
    \midrule
                  Art  &  13  &   6  &   1  &  6 \\
              Biology  &  23  &  12  &   7  &  4 \\
             Business  &  14  &   6  &   2  &  6 \\
            Chemistry  &  19  &  12  &   5  &  2 \\
     Computer science  &  21  &  10  &   7  &  4 \\
            Economics  &  20  &   6  &   8  &  6 \\
          Engineering  &  23  &  15  &   7  &  1 \\
Environmental science  &  18  &   7  &   8  &  3 \\
            Geography  &  17  &   9  &   6  &  2 \\
              Geology  &  21  &  12  &   8  &  1 \\
              History  &   7  &   5  &   1  &  1 \\
    Materials science  &  24  &  15  &   8  &  1 \\
          Mathematics  &  25  &  13  &   8  &  4 \\
             Medicine  &  26  &  14  &  12  &  0 \\
                Other  &   8  &   6  &   2  &  0 \\
           Philosophy  &  12  &   7  &   5  &  0 \\
              Physics  &  39  &  25  &  10  &  4 \\
    Political science  &  13  &   6  &   6  &  1 \\
           Psychology  &  22  &  11  &   7  &  4 \\
            Sociology  &  20  &  13  &   4  &  3 \\
    \bottomrule
    \end{tabularx}
    \caption{Distribution of overall quality scores for readability, split by field of study. Corresponds to distributions shown in Figure~\ref{fig:eval_fos}.}
    \label{tab:eval_raw_by_fos}
\end{table}

\section{Association between paper features and overall readability}
\label{app:eval_association}

To investigate the possibility of identifying paper extractions with major problems, we fit a Logistic Regression classifier using element specific evaluation results as input features, and whether or not a paper has major problems as the target class for classification. Element specific questions are converted into 43 binary input variables;
for example, the title element is mapped to three binary variables, whether the title is extracted correctly (\texttt{title\_yes}), extracted partially (\texttt{title\_partially}), or extracted incorrectly (\texttt{title\_no}). We collapse the targets into two binary classes, 1 if the paper has major problems, and 0 if it has no major problems or some problems. The classifier is trained using 5-fold cross validation, with balanced class weights, and achieves a mean accuracy of 0.69, and area under the ROC of 0.65. This performance is not particularly notable or good; the labeled training sample is small, and due to the complexity of what makes a document problematic to read, we did not expect there to be a clear way to predict extractions with major problems based on a small number of element-level features. Something we aim to explore more in the future is whether the raw tokens on the PDF or publisher metadata can be leveraged to better predict when our extractive parse has failed.

\begin{table}[h!]
    \centering
    \begin{tabular}{r|ccc|c}
        \toprule
        Class & Precision & Recall & F1-score & Support \\
        \midrule
        No major problems / Some problems & 0.91 & 0.71 & 0.80 & 332 \\
        Major problems & 0.23 & 0.55 & 0.32 & 53 \\
        \bottomrule
    \end{tabular}
    \caption{Precision, recall, and F1-scores for classification. The classifier does not perform well at identifying papers with major problems from element-based features (F1 = 0.32).}
    \label{tab:logreg_majorproblems}
\end{table}

The top 10 predictive features and their logistic regression weights are:

\vspace{4mm}
\centering{
\begin{tabular}{rl}
    Abstract extracted incorrectly: & 0.42 \\
    One table extraction error: & 0.18 \\
    No table caption errors: & 0.15 \\
    One figure extraction error: & 0.15 \\
    One figure caption extraction error: & 0.15 \\
    Bibliography extraction is very bad: & 0.13 \\
    More than one figure extraction error: & 0.13 \\
    Authors extracted incorrectly: & 0.12 \\
    More than one table extraction error: & 0.11 \\
    Authors extracted correctly: & 0.09 \\
\end{tabular}}
\vspace{4mm}

\justifying
The most predictive feature is when abstracts are extracted incorrectly. Given the prevalence of abstracts in various literature databases, abstract quality could be easily assessed through external verification. In other words, if the abstract we extract is different from the abstract found for the same paper in other databases on in the publisher metadata, perhaps we can avoid surfacing this paper. However, the distribution of weights among various other element-level features suggests that this feature alone would be insufficient, and that the contributions of these various features are complex, denying us an easy way of identifying paper parses with major problems.

\section{User study materials}
\label{app:user_study_supplement}

Documents used for the user study are provided in this Appendix.

\subsection{Recruitment email}
\label{app:recruitment_email}

The following email was sent and forwarded to several mailing lists to recruit participants.

\begin{center}
\noindent \rule{0.9\linewidth}{0.4pt}
\end{center}
\begin{quote}
The \semanticscholar Research Team at the \allenai is conducting an experiment to evaluate the screen reader accessibility of scientific papers.

\vspace{4pt}\noindent We are looking for participants who are age 18 or older, who identify as blind or low vision, and who have experience using screen readers to interact with scientific papers. If you are interested in participating, please complete the following form to determine eligibility: \underline{link}

\vspace{4pt}\noindent Participation in this study is entirely voluntary. If you do decide to participate, your individual data will be kept strictly confidential and will be stored without personal identifiers.

\vspace{4pt}\noindent The study involves an informational interview to better understand screen reader needs around scientific papers. Each participant will also be asked to interact with papers on a web interface developed by the team. The study will take approximately 75 minutes, and participants will receive a \$150 Amazon gift card for their time.

\vspace{4pt}\noindent Location: Online (Zoom)

\vspace{4pt}\noindent Please contact the authors if you have any questions or concerns about this study. Thank you in advance for your time! Please help us spread the word by forwarding as appropriate.
\end{quote}
\begin{center}
\noindent \rule{0.9\linewidth}{0.4pt}
\end{center}

\subsection{Pre-interview questionnaire}
\label{app:pre_interview_questionnaire}

Prior to each user study interview, the participant was asked to complete the following form:

\begin{center}
\noindent \rule{0.9\linewidth}{0.4pt}
\end{center}
\begin{quote}
\textbf{Share 3 to 5 scientific papers that are difficult to read due to accessibility issues}

\vspace{4pt}\noindent Thank you for volunteering to take part in this study! Please take a few minutes to supply us with some subject keywords you are interested in, and a list of 3 to 5 scientific papers you have found difficult to read due to accessibility issues. This would help us better plan the study based on your experience.

\vspace{4pt}
\begin{enumerate}
    \item[1.] Your name (First name, last initial)
    \item[2.] Please give a few examples of subject keywords you care about. \\ \small For example, computing hardware, analog computer, etc. \normalsize
    \item[3.] Share one paper you have had difficulty reading due to accessibility issues by answering the following questions.
    \begin{itemize}
        \item[--] Paper title \& link \\ \small For example, ``What every Researcher should know about Searching - Clarified Concepts, Search Advice, and an Agenda to improve Finding in Academia'' (https://pubmed.ncbi.nlm.nih.gov/33031639/) \normalsize
        \item[--] On a scale of 1 to 5, how easy or difficult was it for you to read this paper? \\ \small (1 = very easy, 5 = very difficult) \normalsize
        \item[--] Briefly describe why you chose the rating
    \end{itemize}
    \item[4--7.] \textit{Repeat 3}.
\end{enumerate}
\end{quote}
\begin{center}
\noindent \rule{0.9\linewidth}{0.4pt}
\end{center}

\subsection{Interview questions}
\label{app:interview_questions}

The following discussion guide is used to provide structure for user interviews. \\

\textbf{Phase I -- Warmup}:
\begin{itemize}
    \item Can you tell us a little bit about yourself? (Background, what kind of research do you do)
    \item Tell us about how you normally read papers - What is your workflow like? What tools do you use? 
    \begin{itemize}
        \item Do you usually read PDFs directly or do you read papers in other ways?
    \end{itemize}
    \item If you need to read a paper and it is not accessible, what do you do now?
    \begin{itemize}
        \item How long does the process take?
        \item How often is it successful?
    \end{itemize}
    \item Can you give a few examples of the main challenges you face when reading papers?  (For example, are there certain features or attributes of papers that make them particularly difficult to read?) 
    \item In your opinion, are there any resources that provide papers that are easier to read by screen readers? (For example, any journals, conferences, or search engines?)
    \item Overall, how do you feel about your current experience of reading papers? 
\end{itemize}

\textbf{Phase I -- Current workflow}:
\begin{itemize}
    \item Based on the list of papers you provided, walk us through how you would read the paper \texttt{[abc]}. Use the screen reader of your choice, and any additional tools or extensions that are part of your usual process.
    \item Instructions: Please share your whole screen, think aloud and walk me through your thinking process
    \item What kind of information were you looking for, and how did you explore the page to find the information?
    \item On a scale of 1 to 5, how easy or difficult was it to read this paper with the tools, and why? (1 = very easy, 5 = very difficult)
    \item If you could change anything, how could this best meet your needs?
\end{itemize}

\textbf{Phase II}:
\begin{itemize}
    \item We are currently working on an experimental prototype to make papers more easily read by screen readers. Please take a minute to read the about page first: \underline{link}
    \item Based on the list of papers you provided, walk us through how you would read the paper \texttt{[abc]} using this HTML render. You can also use the screen reader of your choice, and any additional tools or extensions that are part of your usual process.
    \item Instructions: 
    \begin{itemize}
        \item We are working with prototypes so not everything works
        \item Please think aloud and walk me through your thinking process
        \item Feel free to provide as many feedback as you can, good or bad
    \end{itemize}
    \item Please take a few more minutes to explore the other parts of this prototype. (e.g. References)
    \item On a scale of 1 to 5, how easy or difficult was it to read this paper with the tools, and why? (1 = very easy, 5 = very difficult)
\end{itemize}

\textbf{Phase III}:
\begin{itemize}
    \item On a scale of 1 to 5, how likely are you to use the HTML render, if it is available to you in the future? (1 = very unlikely, 5 = very likely)
    \item Which features do you consider to be most helpful?
    \item Is there anything it would need to have, or change to convince you to use it?
    \item How do you envision yourself using this tool? How might it fit into your workflow? (For example, would it be an additional extension that is part of your usual process?) 
    \item If you could search for papers and view them in this format, what do you think? 
    \item If you could upload any PDF and create an HTML page like this, what do you think? (Would that be helpful for you, or something you might use, why or why not?)
    \item Are you aware of any other tools that display papers in any way besides PDF?
    \item Do you have any additional feedback about the HTML render or anything else that you would like to share?
    \item Thank you
\end{itemize}

\end{document}